\documentclass[a4paper,fleqn]{cas-dc}

\usepackage[authoryear]{natbib}

\def\tsc#1{\csdef{#1}{\textsc{\lowercase{#1}}\xspace}}
\tsc{WGM}
\tsc{QE}

\usepackage{booktabs}
\usepackage{tabularx}

\usepackage{multirow}

\begin{document}
\let\WriteBookmarks\relax
\def\floatpagepagefraction{1}
\def\textpagefraction{.001}

\shorttitle{Tail Dependence in Cryptocurrency Markets}

\shortauthors{Mallela and Leonelli}

\title[mode=title]{Crashing Together, Rallying Apart: Dynamic Conditional Tail Dependence in Cryptocurrency Markets}

\tnotemark[1]
\tnotetext[1]{This work was supported by the Spanish Agencia Estatal de Investigaci\'{o}n grant PID2023-153222OB-I00.}

\author[1]{Rama Siva Sarwari Mallela}
\ead{rmallela.ieu2021@student.ie.edu}
\credit{Conceptualization, Methodology, Software, Formal analysis, Data curation}

\affiliation[1]{organization={School of Science and Technology, IE University},
            city={Madrid},
            country={Spain}}

\author[1]{Manuele Leonelli}
\cormark[1]
\fnmark[1]
\ead{manuele.leonelli@ie.edu}
\ead[url]{https://manueleleonelli.github.io/}
\credit{Conceptualization, Methodology, Supervision, Validation, Writing}

\cortext[1]{Corresponding author}
\fntext[1]{0000-0002-2562-5192}

\begin{abstract}
Cryptocurrency markets are prone to violent, synchronised drawdowns, challenging the claim that a basket of crypto-assets offers genuine internal diversification. Because standard covariance-based metrics fail to capture asymptotic tail dependence, they systematically understate systemic risk and overstate diversification benefits precisely when markets crash. This study maps the conditional dependence structure of the cryptocurrency market directly in the joint tails, isolating direct extremal linkages from those mediated by the rest of the system. We analyse the daily returns of the thirteen largest cryptocurrencies over a sequence of 89 overlapping windows spanning late 2021 to 2025. We apply dynamic H\"{u}sler-Reiss graphical models of extremes, estimated separately for joint crashes and rallies, and benchmark them against a Gaussian graphical model of ordinary co-movement. The results reveal a near-complete and stable lower-tail graph, an upper tail that thins over time to re-form sectoral structures, and the dissolution of ordinary token categories into a single block anchored by a Bitcoin-Ethereum core. These findings imply that intra-crypto diversification fails on the downside, standard risk models underestimate market-wide crash probabilities by roughly eight-fold, and dynamic extremal graphs offer a superior tool for systemic risk monitoring.
\end{abstract}


\maketitle




\section{Introduction}\label{sec:intro}

Cryptocurrencies have evolved from a niche curiosity into an asset class that institutional and retail investors now hold alongside equities and bonds. The primary financial rationale for holding them rests on diversification: the premise that their returns are sufficiently detached from one another, and from traditional markets, to spread risk. However, the true value of diversification is tested not during ordinary market conditions, but during severe drawdowns. Because the crypto market is young, lightly regulated relative to mature asset classes, and prone to violent, synchronised crashes, understanding the structure of its extreme co-movements is a first-order problem for investors constructing portfolios and regulators monitoring systemic risk.

Whether a crypto portfolio offers genuine diversification depends entirely on dependence in the joint tails. The empirical literature has established two broad regularities: crypto-assets tend to fall together more than they rise together, and systemic distress travels through a core of large-capitalisation coins, chiefly Bitcoin and Ethereum \citep{ahelegbey2021tail,bouri2021quantile}. Yet, two critical questions remain actively contested. First, is the downside asymmetry, the tendency to crash together more than to rally together: a stable property of the market, or an artefact of specific samples? Studies remain sharply divided, with some identifying the lower tail as the stronger conduit of contagion \citep{gkillas2018extreme,naeem2020} and others the upper \citep{naeem2020tail,tiwari2020empirical}. Second, does this extreme dependence structure remain fixed across market cycles, or does it shift dynamically, with evidence split between time-invariant tails \citep{jlassi2023subsample} and continuous structural change \citep{de2020tracking,gong2022asymmetric}.

These questions have proved difficult to settle because standard econometric tools leave crucial dimensions of the joint tail unmeasured. Copula and traditional tail-dependence methods are largely bivariate or low-dimensional; they impose a parametric structure in advance and cannot isolate direct dependence from co-movement mediated by the rest of the market \citep{gkillas2018extreme,tzagkarakis2023}. Quantile connectedness and co-occurrence networks scale to higher dimensions, but they build edges from spillover volumes or joint exceedance counts, meaning a missing edge carries no formal probabilistic interpretation \citep{ahelegbey2021tail,bouri2021quantile}. Most recently, partial-correlation networks have been used to condition on the broader system \citep{naifar2025}, but these rely on second-moment, linear dependence, which coincides with ordinary market conditions and is not specific to the extremes. What is missing is a framework capable of recovering the conditional independence structure of the market in the joint tail itself.

This paper closes that gap. We apply H\"{u}sler-Reiss (HR) graphical models of extremes \citep{engelke2020graphical}, the tail-specific analogue of the Gaussian graphical model (GGM), to the thirteen largest and longest-traded cryptocurrencies. This approach allows us to recover the conditional dependence structure of the joint tail, where an edge between two assets indicates they remain directly dependent at the extremes even after conditioning on all others. To capture structural evolution, we estimate these networks dynamically over 89 overlapping windows spanning late 2021 to 2025. We estimate the graphs separately for the lower tail (joint crashes) and upper tail (joint rallies), and benchmark them against a GGM representing ordinary co-movement on the exact same data. The assets are further grouped into functional categories (currency, smart-contract, payment, and infrastructure) to trace how dependence interacts with token utility.

This is the first application of graphical models of extremes to the cryptocurrency market, and the first to estimate them dynamically against a Gaussian benchmark. Two central themes emerge. First, diversification within the asset class very nearly disappears at the extremes. The lower-tail graph is near-complete and highly stable across the sample, meaning the market crashes as a single block. The upper tail, by contrast, thins over time and begins to re-form a sectoral structure. The crash-rally asymmetry is therefore not a fixed property but an emergent one, a finding that reconciles the conflicting bivariate literature as static estimates of a moving target. Second, the market maintains a rigid hierarchy: Bitcoin and Ethereum form the strongest conditional pair in both tails, while the token categories that organise ordinary co-movement dissolve entirely during crashes. These findings carry a stark warning for risk management: standard covariance-based tools understate the probability of a market-wide crash, offering an illusion of diversification precisely when it is most needed.

The paper is organised as follows. Section~\ref{sec:lit} reviews the related literature. Section~\ref{sec:methods} describes the data and the modelling framework. Section~\ref{sec:results} reports the results. Section~\ref{sec:discussion} discusses their implications and limitations. Section~\ref{sec:conclusion} concludes.

\section{Literature Review}\label{sec:lit}

The empirical literature on cryptocurrency extremes has evolved rapidly, driven by the need to understand whether the asset class offers genuine diversification or harbours concentrated systemic risk. This section reviews the evidence across five main dimensions: the heavy-tailed downside risk of individual coins (Section~\ref{sec:21}), the asymmetry between crash and rally contagion (Section~\ref{sec:22}), the transmission of systemic spillovers (Section~\ref{sec:23}), the role of jumps and bubbles in market structure (Section~\ref{sec:24}), and the structural stability of these linkages over time (Section~\ref{sec:25}). We then survey the modelling landscape (Section~\ref{sec:26}), highlighting the methodological constraints of existing approaches and positioning graphical models of extremes as the natural solution to the gaps they leave open.

\subsection{Extreme risk in individual crypto-assets}\label{sec:21}

The starting point for understanding crypto-market dependence is the recognition that individual cryptocurrency returns carry downside risk far beyond that of conventional assets. Early applications of extreme value theory to the market established that Bitcoin exhibits volatility several times larger than G10 currencies, with extreme losses tending to arrive in consecutive bursts rather than isolated shocks \citep{osterrieder2016}. This heavy-tailed behaviour extends across the broader asset class; returns consistently fall into the Fréchet domain, characterised by power-law tails for which the generalised Pareto distribution is the natural description \citep{gkillas2018application}. In fact, the tails of some major coins are so heavy that they approach or enter an infinite-variance regime \citep{tzagkarakis2023}. 

For portfolio management, this structural feature renders standard covariance-based tools, such as mean-variance optimisation and market beta, highly unreliable. The asset-level evidence points to a clear consensus: because the risk of a single coin is heavy-tailed and clustered, any portfolio or systemic-risk calculation built on a standard covariance matrix will systematically understate the probability of large joint losses. This fundamental limitation highlights the inadequacy of average-dependence metrics and motivates the shift toward explicitly extremal tools to capture market dynamics under stress.
\subsection{Asymmetric contagion: crashes versus rallies}\label{sec:22}

For diversification, the decisive question is not how risky a single coin is, but whether coins fall together more than they rise together, since co-crashes are what erode the benefit of holding several assets at once. On this point, the literature is sharply divided. One body of evidence finds downside dependence dominant, meaning diversification weakens most precisely in market crashes \citep{gkillas2018extreme,naeem2020,liu2025measurement}. Conversely, another strand finds upper-tail dependence to be stronger and more prevalent, making contagion primarily a bull-market phenomenon \citep{tiwari2020empirical,naeem2020tail}, while high-frequency analyses suggest the link can actually weaken as one moves deeper into the extreme tails \citep{chan2022extreme}. The most informative contribution to this debate does not take a static side but explains why the sides differ: \citet{gong2022asymmetric} estimate which dependence regime holds and when, finding that the Bitcoin-Ethereum lower tail has shifted toward a more tightly linked state since 2017.

Despite these insights, two methodological features of this literature limit its usefulness for portfolio-level questions. First, these studies are largely bivariate. Second, they pre-commit to a parametric tail family, forcing the data to choose parameters within an assumed form rather than allowing the true dependence structure to emerge. The persistent disagreement over which tail dominates is itself the crucial lesson: the observed asymmetry depends heavily on the chosen copula family, the conditioning set, the specific coins covered, and the sample period. This implies that the relative strength of crash versus rally co-movement should not be treated as a fixed parameter, but rather as a dynamic, system-wide property that must be estimated and tracked over time.

\subsection{Systemic risk and spillovers}\label{sec:23}

A small set of large cryptocurrencies consistently emerges as the systemic core of the market, even as the identity and ranking of individual hubs shift with the method used. Connectedness in the tails is higher than at the mean, asymmetric between the two tails, and time-varying \citep{bouri2021quantile,waltz2022vulnerability}. Bitcoin and Ethereum recur as the central assets, but their specific roles remain contested: one framework casts them as net transmitters \citep{bouri2021quantile}, another finds Litecoin rather than Ethereum to be the strongest driver of Bitcoin's conditional risk \citep{waltz2022vulnerability}, and a tail-risk network casts Bitcoin as a giver and Ethereum as a receiver of contagion \citep{ahelegbey2021tail}. Beyond the asset class itself, downside spillovers run mainly from equities into crypto rather than the reverse \citep{hanif2022nonlinear}. 

What unites these studies, and limits them for our purpose, is that they quantify transmission and aggregate exposure. They successfully identify hubs and net flows, but they cannot determine whether two coins fall together because one directly moves the other, or because both are responding to a common driver. Filtering out these confounding effects to separate direct dependence from co-movement mediated by the rest of the market requires a conditional-independence structure. This is precisely what existing spillover measures leave unaddressed, and what our graphical network approach provides.

\subsection{Jumps, bubbles and market structure}\label{sec:24}

Extreme moves in crypto markets are frequently discontinuous and partly speculative. Studies of their propagation converge on a clear picture: Bitcoin is the dominant centre through which extreme events spread, while the precise shape of the network depends on the kind of event. Jump-based analyses show that simultaneous extreme moves amplify future risk and that different jump types act in opposite directions \citep{gkillas2024discontinuous,gkillas2025heterogeneity}. Network studies of co-jumps and co-bubbles confirm Bitcoin's central role, while finding that co-jump transmission follows a market-capitalisation hierarchy whereas co-bubble transmission is more diffuse beyond the largest coins \citep{zhang2023co,chen2024co}. From a microstructure angle, \citet{desagre2023crypto} trace how liquidity behaves around extreme returns and document a maturation of the market from liquidity-driven toward trading-driven dynamics. 

As with the systemic-risk measures discussed above, the edges in these networks are defined by the co-occurrence of extreme events rather than by a formal model of conditional dependence.

\subsection{Structural change in extremal dependence}\label{sec:25}

Whether extreme dependence is a fixed feature of the market or one that evolves with the cycle determines whether a single static estimate can suffice. Here, too, the evidence divides. \citet{jlassi2023subsample} find the stock-crypto tail copula to be time-invariant, concluding that extreme dependence is structural rather than crisis-induced, even as mean correlations move with crises. Conversely, \citet{gong2022asymmetric}, as noted above, find continuous evolution in the dependence itself, and \citet{de2020tracking} identify a discrete break datable to the onset of the 2018 crash using a non-parametric extreme-value method. 

These readings are complementary rather than contradictory, since the form of the dependence may be stable across subperiods while its strength shifts. Together, they demonstrate that the temporal behaviour of extremal dependence cannot be assumed away, and is best examined as the market moves through successive cycles. This is precisely the dynamic our windowed design is built to address for the internal structure of the crypto market.

\subsection{Modelling landscape}\label{sec:26}

The approaches above can be read as a progression in how much of the joint tail they recover, summarised in Table~\ref{tab:landscape}. Across this literature, two limitations recur. First, copula and tail-dependence methods are largely low-dimensional and typically fix a parametric tail family in advance \citep{tiwari2020empirical,trucios2020value}; even semiparametric treatments that relax the tail-family assumption remain, for tractability, essentially bivariate \citep{leonelli2020semiparametric}. They describe the joint tail within an assumed form rather than letting its structure emerge, and they cannot separate direct dependence from co-movement mediated by other assets. Second, network and connectedness methods build edges from the co-occurrence of extreme events or regression sensitivities, meaning a missing edge carries no probabilistic interpretation. The most recent and closest design, \citet{naifar2025}, builds partial-correlation networks across token groups for each tail and likewise finds downside dependence to be stronger. However, while partial-correlation networks do account for indirect relationships, they rest on second-moment, linear dependence. They do not provide a dedicated representation of extremal dependence.

A recent alternative removes both limitations at once. Building on \citet{engelke2020graphical}, the graphical-extremes framework places a graph directly on the limiting distribution of the joint exceedances. In this framework, a missing edge means two assets are conditionally independent in the extremes once the others are accounted for. This provides the formal probabilistic statement that co-occurrence networks lack, and it isolates the tail itself rather than the centre of the distribution where partial-correlation methods operate. 

This distinction naturally motivates our empirical strategy. The Gaussian benchmark we adopt describes the conditional dependence of ordinary conditions in the bulk of the distribution. The extremal graph, by contrast, describes the dependence specific to the tail. The contrast between the two reveals exactly how the market's structure under stress departs from its structure in calm periods. Furthermore, because the graph is read off an estimated dependence model, it can be recovered separately for the lower and upper tails and re-estimated through time, turning a static snapshot into a dynamic trajectory. The framework has begun to see use in finance, applied to asset returns by \citet{kluppelberg2021estimating} and to foreign exchange rates by \citet{engelke2022structure}. To our knowledge, however, it has not previously been brought to the cryptocurrency market, nor estimated dynamically, which is the contribution of the present paper.

\begin{table*}
\centering
\small
\setlength{\tabcolsep}{4pt}
\renewcommand{\arraystretch}{1.25}
\caption{Approaches to dependence in cryptocurrency markets and what they leave unmeasured
at the extremes.}
\label{tab:landscape}
\begin{tabularx}{\textwidth}{@{}p{3cm} p{3cm} p{3cm} X@{}}
\toprule
Approach & Representative work & Captures & What it leaves unmeasured at the extremes \\
\midrule
Univariate extreme-value analysis &
\citep{osterrieder2016,gkillas2018application} &
Tail risk of a single asset &
Any co-movement between assets \\
Pairwise and vine copulas &
\citep{gkillas2018extreme,tiwari2020empirical,trucios2020value} &
Tail dependence of asset pairs &
High-dimensional structure; assumes a tail family; cannot separate direct from mediated
dependence \\
Quantile connectedness and CoVaR &
\citep{bouri2021quantile,waltz2022vulnerability} &
Directed spillovers and systemic exposure &
Which dependences are direct rather than transmitted through other assets \\
Co-jump, co-bubble and tail-risk networks &
\citep{ahelegbey2021tail,zhang2023co} &
A graph of extreme co-movement &
Edges are co-occurrence counts, not probabilistic conditional independence \\
Partial-correlation tail networks &
\citep{naifar2025} &
A conditional, tail-specific graph &
Edges rest on second-moment dependence, not dependence in the tail \\
\midrule
\textbf{Graphical models for extremes (this paper)} &
\citep{engelke2020graphical} &
Conditional independence in the joint tail &
Resolves the above: missing edges are conditional independence in the extremes, by tail and
over time \\
\bottomrule
\end{tabularx}
\end{table*}

\subsection{Summary and outlook}\label{sec:27}

The literature documents the heavy downside risk of individual coins, the systemic-risk role of a Bitcoin and Ethereum core, and the unresolved debate over crash-versus-rally contagion. Across these strands, extreme co-movement consistently exceeds average co-movement \citep{ahelegbey2021tail,bouri2021quantile,gong2022asymmetric}. Consequently, risk and diversification measures built on standard covariance matrices systematically understate joint tail behaviour. 

By applying the graphical-extremes framework outlined above, this paper directly settles the two open empirical questions: whether the crash-rally asymmetry is a stable property or a moving target, and whether the market's core-periphery structure holds across cycles. For an investor, resolving this identifies exactly when the diversification apparent in normal conditions disappears, which is critical when rebalancing a crypto-inclusive portfolio across changing market regimes. For a regulator, it maps how systemic linkage concentrates and shifts across token categories, providing the foundation for tail-sensitive market monitoring. Ultimately, this approach bridges the theoretical rigour of extreme-value theory with the practical assessment of diversification and systemic risk.


\section{Data and Methodology}\label{sec:methods}

\subsection{Data}\label{sec:data}

We study the thirteen cryptocurrencies listed in Table~\ref{tab:coins}, selected by market capitalisation and by length of available history, so that the panel covers the largest and most economically significant assets while sharing a common sample long enough to support a windowed analysis. We collect daily adjusted closing prices from 1 October 2020 to 14 June 2026 and convert them to logarithmic returns. For the dependence and network analysis the assets are assigned to four functional groups that reflect their economic role: currencies, used primarily as a store of value or medium of exchange; smart-contract platforms, which host decentralised applications; payment tokens, oriented toward transfers and settlement; and infrastructure tokens, which provide cross-chain and protocol-level services.

\begin{table}[t]
\centering
\small
\caption{Cryptocurrencies in the sample, grouped by functional category.}
\label{tab:coins}
\begin{tabular}{@{}lll@{}}
\toprule
Category & Cryptocurrency & Symbol \\
\midrule
Currency & Bitcoin & BTC \\
 & Litecoin & LTC \\
 & Dogecoin & DOGE \\
\midrule
Smart-contract platform & Ethereum & ETH \\
 & Solana & SOL \\
 & Cardano & ADA \\
 & Avalanche & AVAX \\
\midrule
Payment & XRP & XRP \\
 & Stellar & XLM \\
 & Tron & TRX \\
\midrule
Infrastructure & BNB & BNB \\
 & Chainlink & LINK \\
 & Polkadot & DOT \\
\bottomrule
\end{tabular}
\end{table}

Because the windowed design requires a common history across the whole sample, assets that did not survive the period are excluded by construction, most notably the Terra ecosystem tokens that collapsed in 2022. Stablecoins are likewise excluded; their pegged nature and lack of meaningful continuous variation violate the assumptions of the marginal filtering required to isolate dependence. The panel is therefore conditioned on survival and floating exchange rates, as is usual in analyses that require a balanced set of series, and the dependence structure we recover is that of the assets which persisted through the sample.

\subsection{Marginal filtering and diagnostics}\label{sec:filtering}

Because simultaneous extreme co-movement is the object of interest, we first strip out the marginal dynamics that would otherwise contaminate a dependence analysis. Each return series is filtered with a first-order autoregressive mean and an asymmetric GJR-GARCH$(1,1)$ variance with Student-$t$ innovations \citep{glosten1993relation}. This specification is chosen because it explicitly accounts for the leverage effect, heavy tails, and short-lived mean autocorrelation characteristic of cryptocurrency returns. The analysis is then conducted on the resulting standardised residuals $\mathbf{Z} = (z_{i,t})$, where $z_{i,t}$ is the residual of asset $i$ at time $t$. This two-step separation of marginal dynamics from dependence follows \citet{mcneil2000estimation}. 

We confirm the adequacy of the filtering on every series. As detailed in Appendix~\ref{app:diagnostics}, a full battery of diagnostic tests confirms that the filter successfully removes serial correlation and volatility clustering across the panel. Crucially, the standardised residuals remain strongly leptokurtic, retaining finite variance but heavy, sub-quartic tails. This is precisely the marginal behaviour that a covariance-based view of dependence would misrepresent and that an extremal model is designed to exploit, motivating the two complementary dependence models that follow.

For the Gaussian baseline, the standardised residuals are used directly. For the extremal analysis, each series is transformed to standard Pareto margins,
\begin{equation}
x_{i,t} \;=\; \frac{1}{1 - \hat{F}_i(z_{i,t})},
\end{equation}
where $\hat{F}_i$ is the empirical marginal distribution of $z_{i,t}$. We define the extreme tail using a threshold of $p = 0.20$: if any asset's residual exceeds its 80th percentile on a given day, the entire multivariate observation is retained. This places every series on a common scale so that joint tail behaviour is not confounded by differences in the marginals, a standard routine in graphical extremes \citep{engelke2020graphical}. Appendix~\ref{app:robustness} confirms that our structural findings are highly robust to alternative thresholds ($p = 0.10$ and $0.15$). To study simultaneous crashes, the same transformation is applied to $-z_{i,t}$, so that the lower tail of returns becomes the upper tail of the transformed series. Throughout, positive and negative extremes are estimated as separate networks, allowing the possibility that joint surges and joint crashes propagate through structurally different dependence structures.

\subsection{Gaussian graphical models}\label{sec:ggm}

As a baseline characterising dependence in ordinary conditions, we estimate a GGM \citep{lauritzen2026graphical}. Suppose the standardised residual vector $\mathbf{Z} = (Z_1, \ldots, Z_d)$ follows a multivariate Gaussian distribution with precision matrix $\Theta = \Sigma^{-1}$, where $\Sigma$ is the covariance matrix. The off-diagonal entries of $\Theta$ encode the partial correlation between $Z_i$ and $Z_j$ controlling for all the remaining assets,
\begin{equation}
\rho_{ij} \;=\; -\,\Theta_{ij} \,/\, \sqrt{\Theta_{ii}\,\Theta_{jj}}.
\end{equation}
A zero partial correlation corresponds to conditional independence between $Z_i$ and $Z_j$ given the rest of the system, equivalently to the absence of an edge between $i$ and $j$ in the graph. The model therefore turns the precision matrix into a network whose edges represent the conditional co-movement that survives once dependence routed through other assets has been removed.

We estimate the model in a Bayesian fashion \citep{mohammadi2015bayesian}, recovering the posterior distribution of the precision matrix under a non-informative prior and including an edge $(i,j)$ whenever the $95\%$ credible interval of the corresponding partial correlation excludes zero. The resulting graph is weighted and signed: a positive edge indicates conditional co-movement, a negative edge conditional opposition. Bayesian estimation avoids the need to select a penalty by cross-validation and provides direct uncertainty quantification on edge inclusion. Furthermore, because graph density is itself sensitive to the estimation method, Appendix~\ref{app:robustness} compares this Bayesian baseline against a frequentist graphical lasso and finds that the density of the Gaussian graph depends heavily on the selection rule. For that reason we do not rest the contrast between ordinary and extreme dependence on graph density, but on the tail-dependence functionals of Section~\ref{sec:econ}, where the Gaussian benchmark fails in a way no estimator can repair.

\subsection{H\"{u}sler-Reiss graphical models for extremes}\label{sec:hr}

Standard graphical models describe conditional dependence at the centre of the joint
distribution and can fail to represent how assets co-move at the extremes. A natural pairwise
measure of extremal dependence is the extremal correlation,
\begin{equation}
\chi_{ij} \;=\; \lim_{u \to 1} P\!\left( F_j(X_j) > u \mid F_i(X_i) > u \right)
\;\in\; [0, 1],
\end{equation}
the limiting probability that one asset is extreme given that another is. A value
$\chi_{ij} = 0$ indicates asymptotic independence, so that even when each asset is extreme on
its own the two do not tend to be extreme together, while $\chi_{ij} > 0$ indicates
asymptotic dependence, so that extreme events tend to co-occur. Because $\chi_{ij}$ is defined
entirely through the joint tail, two assets with strong ordinary correlation can be
asymptotically independent, and conversely.

The HR Pareto distribution plays the same role for extremal dependence that the
multivariate Gaussian plays for ordinary dependence. Whereas the Gaussian describes the
centre of a distribution through a covariance matrix, the HR distribution
describes the joint tail through a symmetric variogram matrix
$\Gamma \in \mathbb{R}^{d \times d}$, whose entries measure pairwise extremal dependence:
small $\Gamma_{ij}$ indicates strong tail co-movement, large $\Gamma_{ij}$ near-independence
in the tail. It is the multivariate extreme-value distribution whose conditional independence
structure can be encoded through a graph in direct analogy with the Gaussian case
\citep{engelke2020graphical,hentschel2025statistical}. The associated extremal graphical model
defines an undirected graph $\mathcal{G} = (V, E)$ on the assets $V = \{1, \ldots, d\}$, in
which an edge between $i$ and $j$ is present if and only if $X_i$ and $X_j$ are conditionally
dependent in the tail given all the remaining assets. The absence of an edge means that the
extremal dependence between $X_i$ and $X_j$ is fully mediated by the rest of the system, the
tail analogue of the conditional independence that underlies the Gaussian model.

The variogram matrix and the graph are recovered together by minimising a penalised
extremal log-likelihood,
\begin{equation}
\hat{\Gamma},\, \hat{\mathcal{G}} \;=\;
\arg\min_{\Gamma,\,\mathcal{G}}
\left\{ -\ell(\Gamma; \mathbf{X}) + \rho\,\|\Gamma\|_1 \right\},
\end{equation}
where $\ell$ is the HR log-likelihood, $\|\Gamma\|_1$ is the off-diagonal
$L^1$ norm, and $\rho > 0$ is a sparsity penalty. This is the extremal analogue of the
graphical lasso \citep{friedman2008sparse}: the penalty shrinks small entries of $\Gamma$ to zero,
and the resulting zero pattern defines the estimated graph. The selection of $\rho$ is tied
to the dynamic design and is described in Section~\ref{sec:windows}. From the fitted
variogram we obtain the model-implied extremal correlation through the HR
mapping
\begin{equation}
\chi_{ij} \;=\; 2 - 2\,\Phi\!\left( \tfrac{1}{2}\sqrt{\Gamma_{ij}} \right),
\end{equation}
where $\Phi$ is the standard normal distribution function. Reading $\chi$ off the fitted
model rather than estimating it directly from tail counts yields smooth and comparable
pairwise dependence trajectories, which we report alongside the graphs.

\subsection{Dynamic estimation over sliding windows}\label{sec:windows}

To trace how the dependence structure evolves, we re-estimate all three networks (the positive-tail extremal graph, the negative-tail extremal graph, and the Gaussian baseline) on overlapping windows of $750$ days, approximately two years, advanced in steps of $15$ days, dating each window by its mid-point. Because cryptocurrencies trade on every calendar day, a $750$-day window spans about two years of data. This yields $89$ windows whose mid-points span late 2021 to 2025, the final window being centred in 2025 even though the price series extend to mid-2026. The window length is long enough to estimate a thirteen-node extremal graph from the upper fifth of the observations, yet short enough to track structural change, and the fifteen-day step traces a smooth trajectory without redundant computation. Appendix~\ref{app:robustness} confirms that our core structural findings are highly robust to alternative window lengths of $500$ and $1000$ days.

To keep successive estimates comparable, the sparsity penalty is not re-tuned in each window. A single penalty is chosen once, separately for the positive and the negative tail, by ten-fold cross-validation maximising the held-out extremal log-likelihood on the full sample, and is then held fixed across all windows. Re-selecting the penalty window by window would let the density of the estimated graphs move with the tuning rather than with the market, which is exactly the variation the dynamic analysis is intended to measure; fixing it ensures that changes in network structure across windows reflect changes in dependence and not changes in regularisation.

\subsection{Network analysis}\label{sec:network}

For each estimated graph we compute a targeted set of measures to summarise its global topology \citep{newman2018networks}. At the global level we report the edge density, the fraction of possible edges that are present; the diameter and average degree, summarising how compact the network is; the global transitivity, indicating whether dependence propagates in tightly knit triangles; and the degree assortativity, indicating core-periphery or hub-and-spoke patterns. The degree of community structure is quantified using Louvain modularity, which measures how far the network separates into distinct groups.

Because dense extremal graphs render standard binary centrality measures largely uninformative, we evaluate the structural hierarchy of individual assets directly through their pairwise edge weights (the extremal correlations) rather than through unweighted node centrality.

\subsection{Software and reproducibility}\label{sec:repro}

All computations were carried out in R. The daily price series are publicly available from
Yahoo Finance. The code
implementing the methods above, together with scripts to reproduce every result and figure in
the paper, is available at \url{https://github.com/manueleleonelli/HR\_crypto}.


 
\section{Results}\label{sec:results}

This section presents the empirical findings. We first characterise the topology of the cryptocurrency market at the extremes, demonstrating that the conditional independence structure of the joint tail is near-complete and structurally distinct from ordinary co-movement (Section~\ref{sec:41}). Next, we trace the dynamic evolution of these networks over the 89 overlapping windows, revealing that the crash-rally asymmetry is an emergent property rather than a fixed one (Section~\ref{sec:42}), and we confirm with a within-window bootstrap that this asymmetry is statistically significant (Section~\ref{sec:asym}). Finally, we quantify the economic implications of these structures, showing how standard covariance-based models systematically understate the probability of market-wide crashes even where they remain serviceable for ordinary portfolio risk (Section~\ref{sec:43}).

\begin{table*}[t]
\centering
\small
\caption{Network statistics averaged over the $89$ windows, for the lower-tail and upper-tail
extremal graphs and the Gaussian benchmark. Edge density is the share of the $78$ possible
edges that are present; transitivity is the global clustering coefficient; modularity and
assortativity are the Louvain modularity and the degree assortativity.}
\label{tab:globalstats}
\begin{tabular}{@{}lrrrrrr@{}}
\toprule
Network & Density & Avg.\ degree & Transitivity & Modularity & Diameter & Assortativity \\
\midrule
Lower tail (extremal) & 0.88 & 10.6 & 0.88 & 0.015 & 2.0 & $-0.14$ \\
Upper tail (extremal) & 0.86 & 10.4 & 0.86 & 0.018 & 2.0 & $-0.17$ \\
Gaussian (benchmark)  & 0.46 &  5.5 & 0.47 & 0.152 & 2.8 & $-0.15$ \\
\bottomrule
\end{tabular}
\end{table*}

\begin{figure*}[t]
\centering
\includegraphics[width=0.8\linewidth]{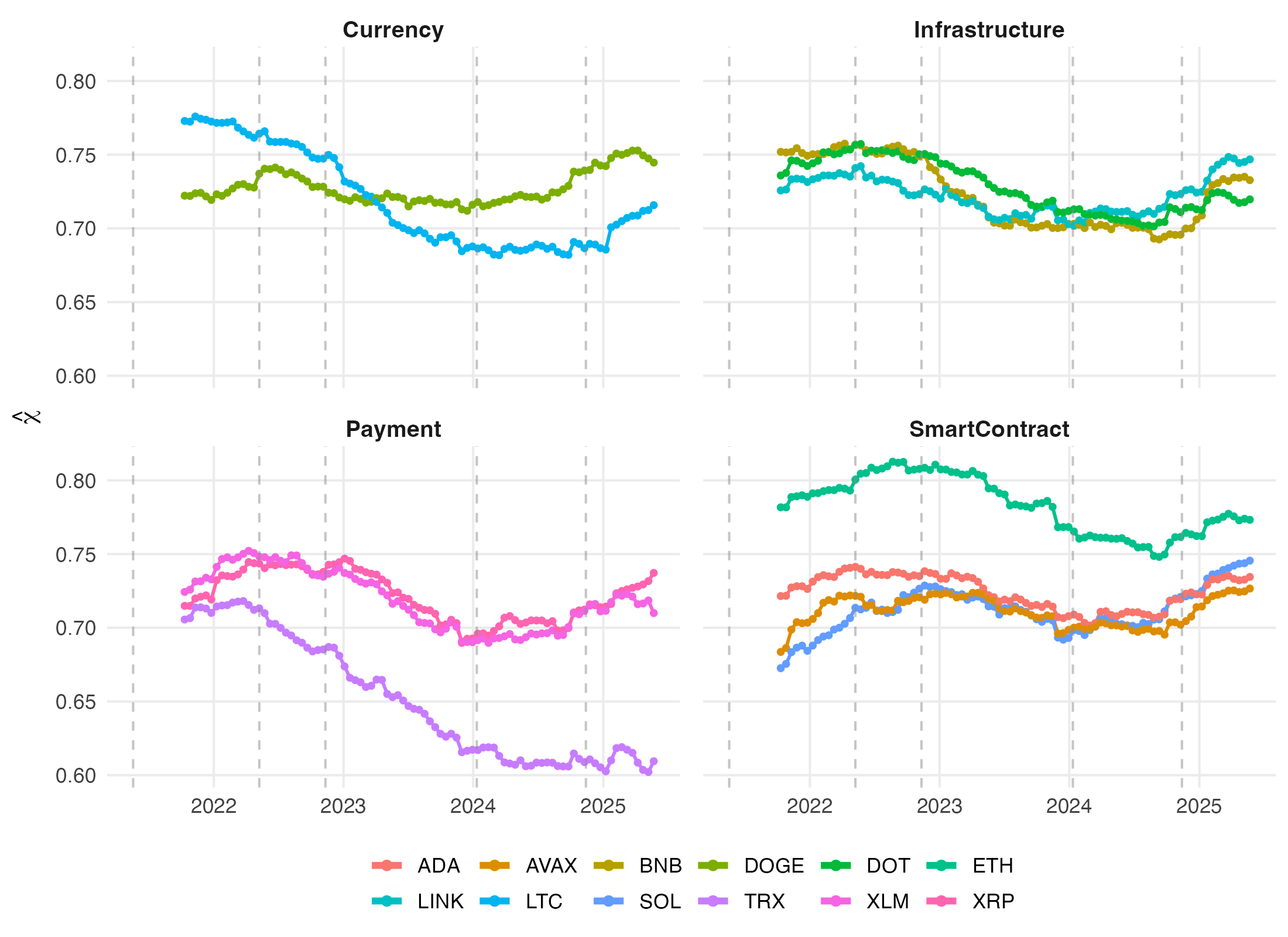}\\
\includegraphics[width=0.8\linewidth]{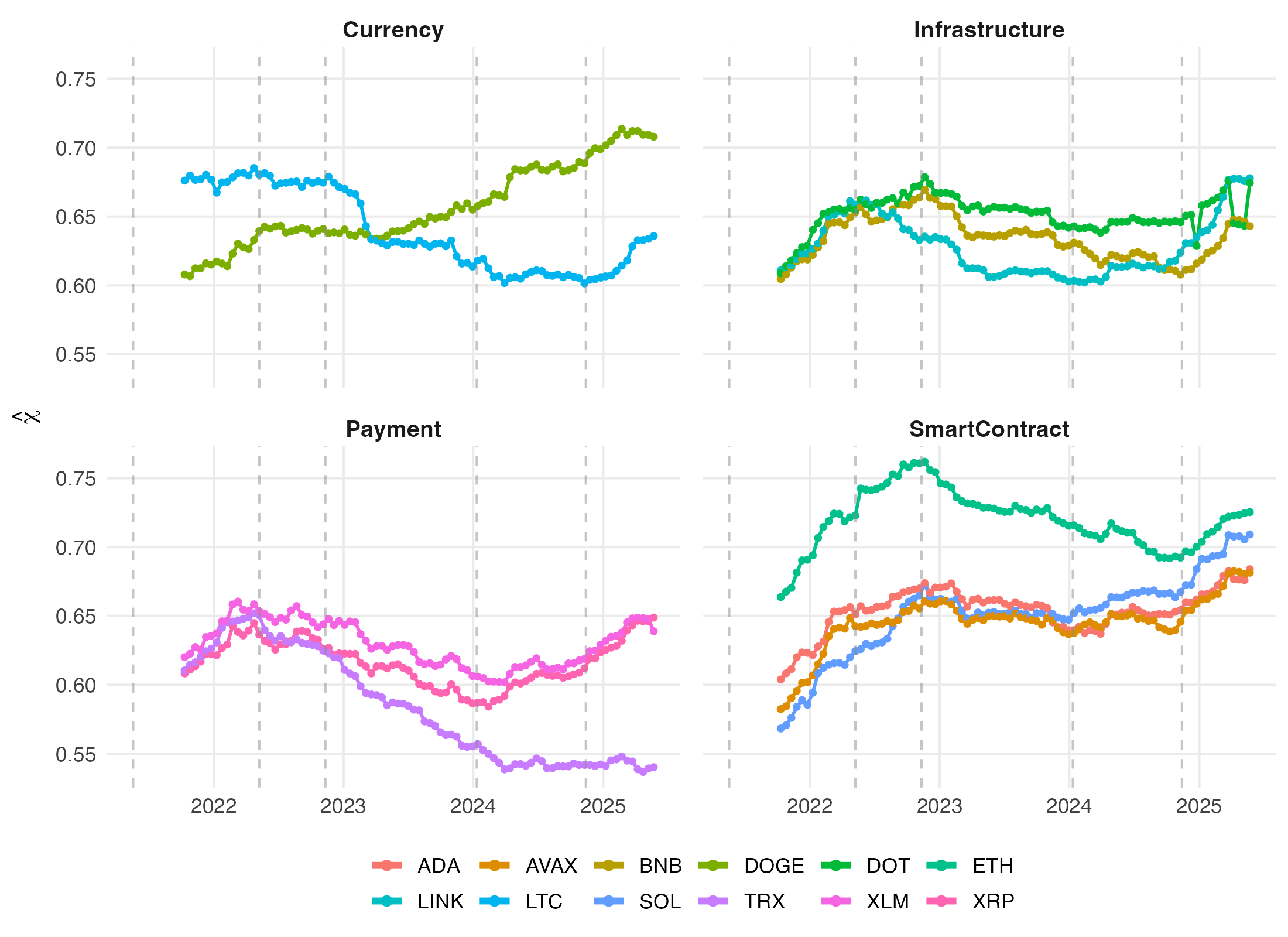}
\caption{Pairwise extremal correlation $\hat{\chi}$ of Bitcoin with each other asset, grouped
by the partner's sector, over the $89$ windows. Top: lower (negative) tail; bottom: upper
(positive) tail. Dashed vertical lines mark the events annotated in
Figure~\ref{fig:temporal}.}
\label{fig:btc_chi}
\end{figure*}

\subsection{The structure of extremal dependence}\label{sec:41}

At the extremes the market is close to fully connected. Table~\ref{tab:globalstats} reports the network statistics averaged over the $89$ windows. Both extremal graphs carry an edge density close to $0.87$; the average asset is directly linked to between ten and eleven of the twelve others in either tail; the global transitivity, the share of connected triples that close into triangles, is near $0.87$; and the diameter is two in every single window, so any two assets are either directly linked or joined through one intermediary. The conditional independences that might organise dependence are, in other words, almost absent in the tails: when one asset moves to an extreme, almost every other asset is liable to follow. The Gaussian benchmark in Table~\ref{tab:globalstats} is sparser and retains a longer diameter, but the edge density of a graphical model is governed by the sparsity-selection rule as much as by the underlying dependence, and a frequentist graphical lasso fitted to the same residuals is denser still (Appendix~\ref{app:robustness}). We therefore do not rest the contrast between ordinary and extreme dependence on density. The substantive differences lie in the topology, to which we now turn, and in the tail-dependence functionals of Section~\ref{sec:econ}, where the Gaussian benchmark fails in a way no choice of penalty can repair.
 
\begin{figure*}[t]
\centering
\includegraphics[width=0.85\linewidth]{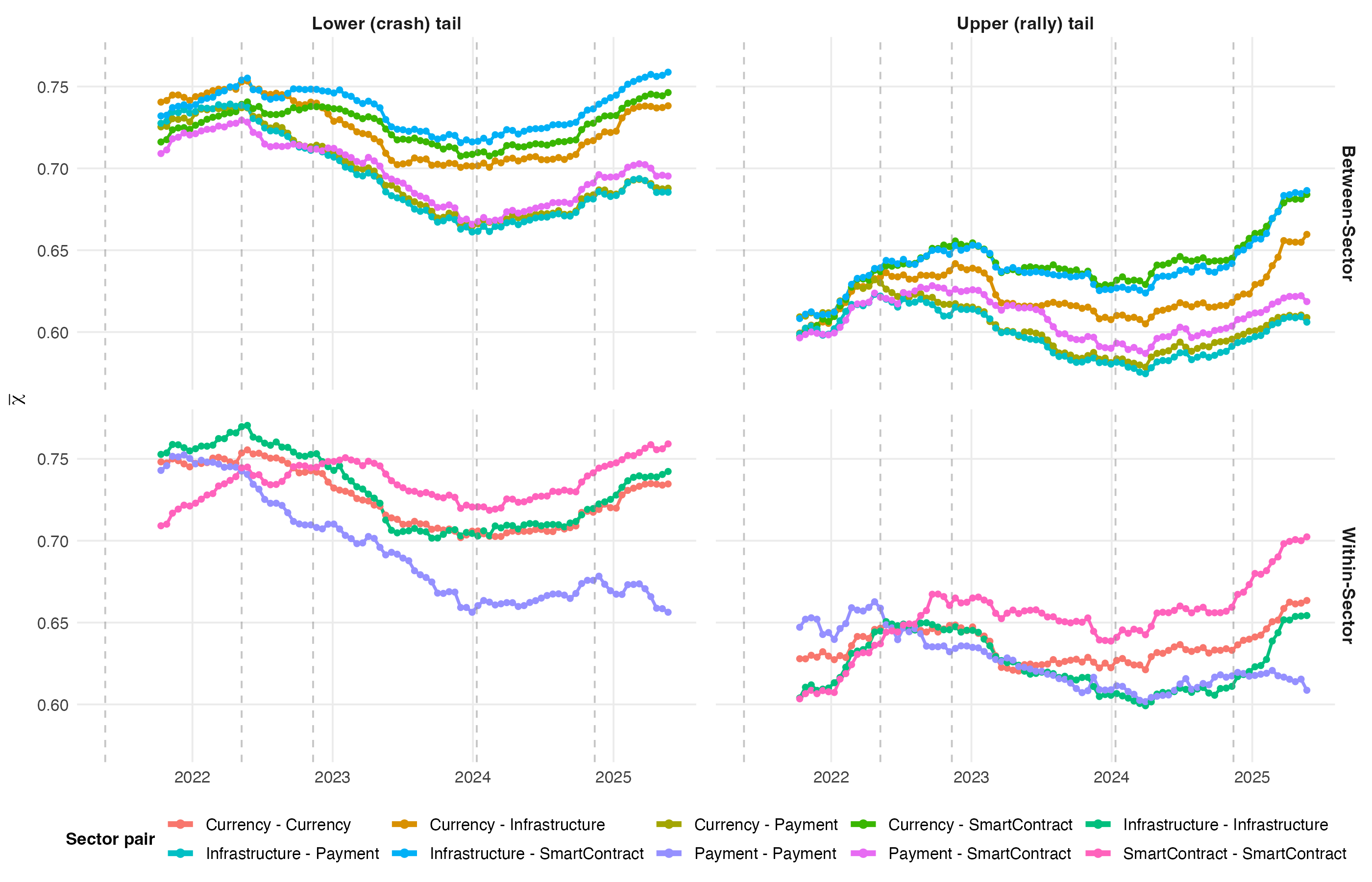}
\caption{Average extremal correlation $\bar{\chi}$ within and between sectors, by tail, over
the $89$ windows. Within-sector pairings (bottom row) and between-sector pairings (top row)
track one another closely, and both lie higher in the lower (negative) tail than in the upper
(positive) tail.}
\label{fig:sector_chi}
\end{figure*}

The same contrast appears in the community structure. The Gaussian graph retains appreciable modularity, around $0.15$, reflecting the sectoral organisation of the assets, whereas both extremal graphs have modularity below $0.02$. The sector-averaged correlations of Figure~\ref{fig:sector_chi} confirm this without reference to any estimator: at the extremes within-sector dependence is no higher than between-sector dependence. Whatever separation between currencies, smart-contract platforms, payment tokens and infrastructure tokens is present in normal-times dependence dissolves at the extremes, where the assets behave as a single tightly connected block.
 
\begin{figure*}[t]
\centering
\includegraphics[width=0.8\linewidth]{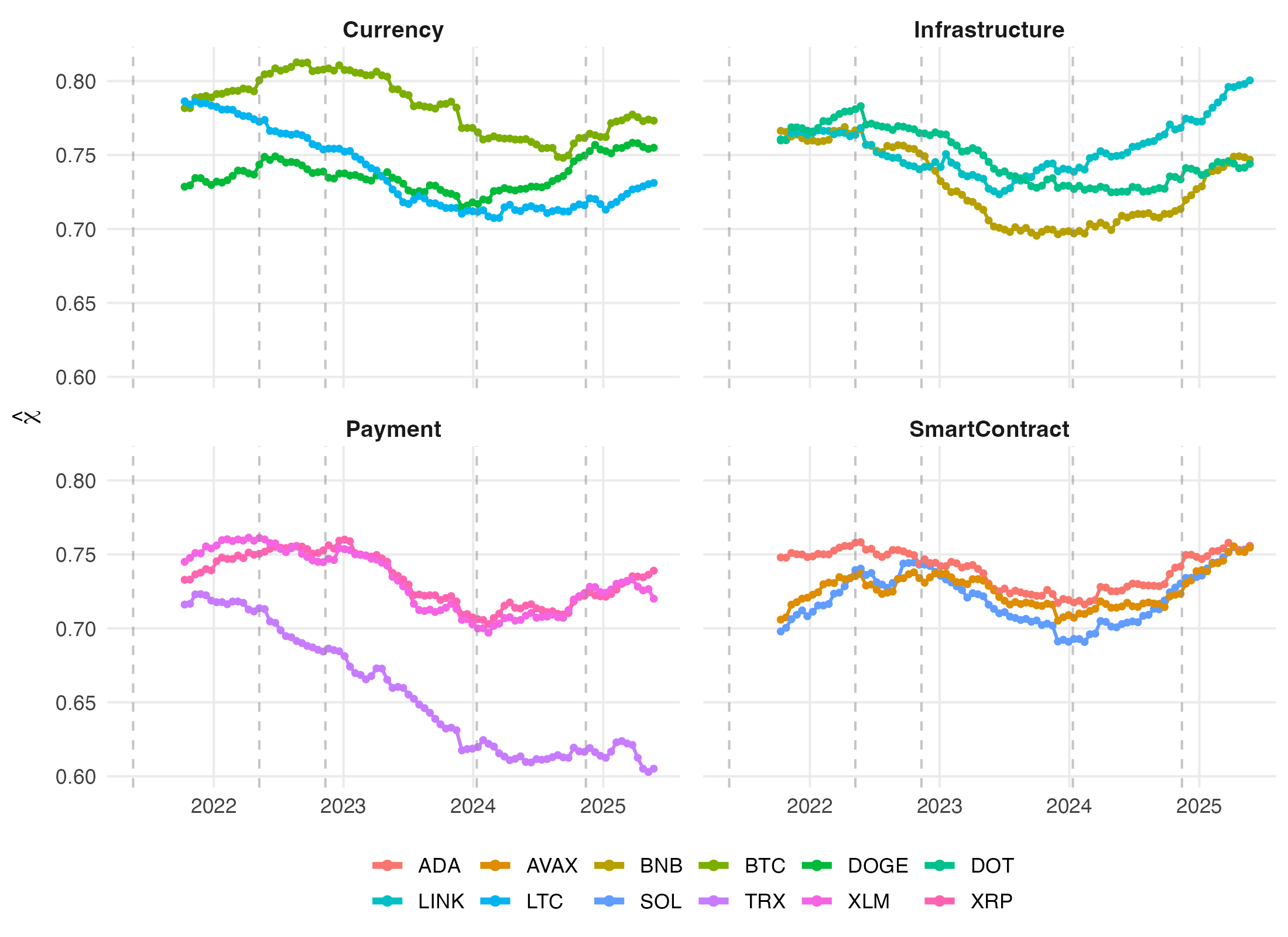}\\
\includegraphics[width=0.8\linewidth]{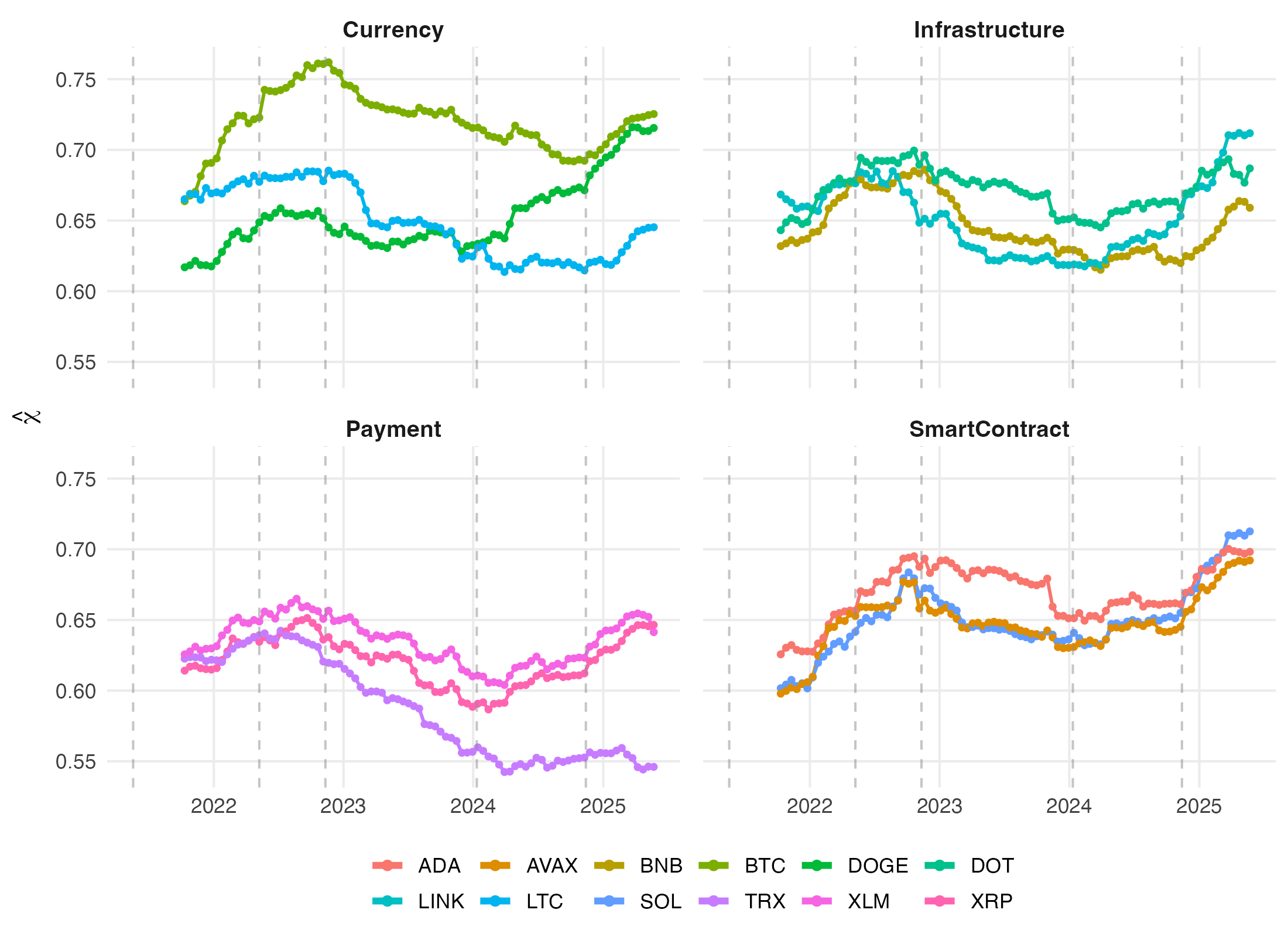}
\caption{Pairwise extremal correlation $\hat{\chi}$ of Ethereum with each other asset, grouped
by the partner's sector, over the $89$ windows. Top: lower (negative) tail; bottom: upper
(positive) tail.}
\label{fig:eth_chi}
\end{figure*}

Beneath this near-complete topology, the strength of pairwise dependence is far from uniform, and it is here that the asset hierarchy appears. Reading this hierarchy directly from the pairwise extremal correlations (Figures~\ref{fig:btc_chi} and~\ref{fig:eth_chi}), Bitcoin and Ethereum form the strongest pair in both tails, with an extremal correlation of $0.78$ in the lower tail and $0.72$ in the upper. Together, they anchor a dense core to which the remaining large-capitalisation assets attach closely. Tron is the consistent periphery: its extremal correlation with both Bitcoin and Ethereum averages about $0.65$ in the lower tail and $0.59$ in the upper, the lowest in the panel in either tail. The asymmetry between the tails is without exception: in every one of the $89$ windows, and for each of the twenty-four pairings of Bitcoin or Ethereum with another asset, dependence is stronger in the lower tail than in the upper, by $0.08$ on average. Joint losses are more tightly linked than joint gains throughout.
 
The sector-averaged extremal correlations describe the same structure at a coarser resolution (Figure~\ref{fig:sector_chi}). Within-sector and between-sector dependence move together and at broadly similar levels, with the smart-contract platforms forming the tightest within-sector group and the payment tokens the weakest, the latter held down by Tron. As with the pairwise correlations, every sector pairing sits higher in the lower tail than in the upper.

 \subsection{Dynamic evolution of the dependence structure}\label{sec:42}

\begin{figure*}[t]
\centering
\includegraphics[width=0.95\linewidth]{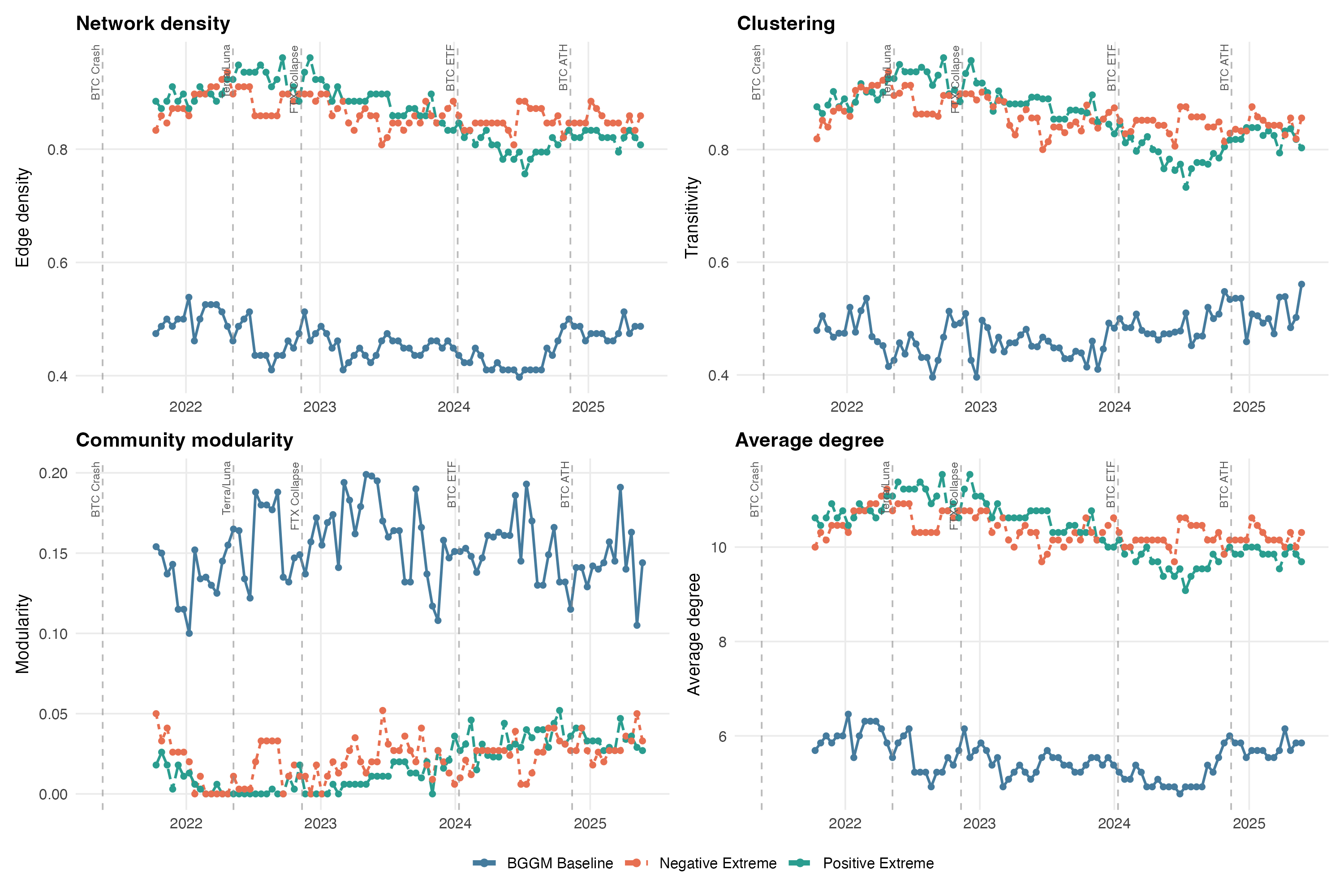}
\caption{Global network statistics over the $89$ windows for the lower-tail extremal graph, the upper-tail extremal graph, and the Gaussian benchmark: edge density, global transitivity, Louvain modularity, and average degree. Dashed vertical lines mark major market events.}
\label{fig:temporal}
\end{figure*}

The window estimates share a common arc (Figure~\ref{fig:temporal}). Extremal dependence is high through 2022, when both tails reach an edge density near $0.94$, declines to a trough in mid-2024 following the approval of spot exchange-traded funds in January, and recovers through the rally that runs into late 2024 and 2025. The pairwise correlations of Bitcoin and Ethereum follow the same path, the lower-tail pair reaching its minimum in September 2024, and the trough is common to the density, the average degree and the transitivity of both extremal graphs. The Gaussian benchmark, by contrast, is stable over the sample, showing neither the trough nor the upper-tail thinning, so the dynamics described here are a property of the tails and not of ordinary co-movement.
 
The asymmetry between the tails, however, is not a fixed feature of the market but one that emerges over the sample, and this is the central dynamic finding. Early in the period the two tails are equally connected, and the upper tail is if anything marginally the denser: in 2021 its average edge density exceeds the lower tail's by $0.03$. The ordering then reverses. Upper-tail density falls from a peak of $0.92$ in 2022 to $0.88$ in 2023 and $0.81$ in 2024, while lower-tail density barely moves, holding between $0.87$ and $0.88$ across the same years. By 2024 the lower tail is the denser of the two by $0.06$, and it remains so in 2025. The same divergence shows in the average degree, which falls to below ten in the upper tail while holding near ten and a half in the lower, and in the modularity, which rises in the upper tail from near zero to about $0.035$ as a faint sectoral structure begins to reassert itself among joint gains, while the lower tail shows no such loosening. Joint losses remain almost universally connected throughout.
 
\begin{figure*}[t]
\centering
\includegraphics[width=0.85\linewidth]{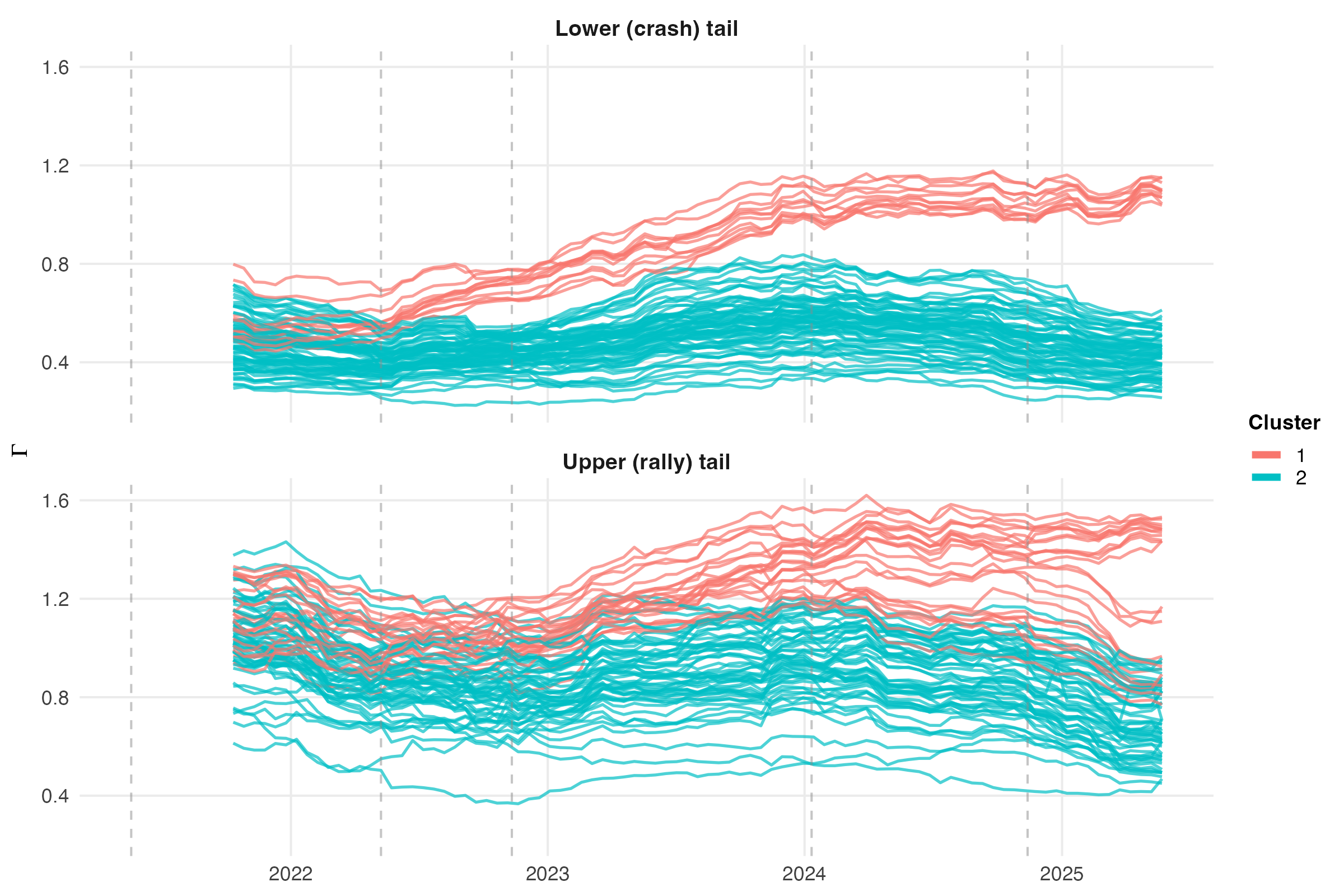}
\caption{Evolution of the pairwise variogram entries $\Gamma_{ij}$ over the $89$ windows, coloured by cluster, for the lower (negative) and upper (positive) tails. A small weakly-dependent cluster (high $\Gamma$) separates from a large strongly-dependent core; it comprises only the twelve Tron pairs in the lower tail and widens to twenty-three pairs in the upper.}
\label{fig:gamma}
\end{figure*}

The variogram clustering corroborates the periphery finding and its asymmetry (Figure~\ref{fig:gamma}). Applying K-means clustering to the pairwise trajectories separates a small group of weakly dependent pairs from a large strongly dependent core. In the lower tail this weak group consists of exactly the twelve pairs involving Tron and nothing else, so the periphery of the crash-time network is a single asset. In the upper tail the weak group widens to twenty-three pairs: it keeps the twelve Tron pairs and adds eleven more, centred on BNB, XRP and Solana, that decouple among the infrastructure, payment and smart-contract tokens. This is the disaggregated counterpart of the upper tail's thinning. It is not merely that fewer edges survive in booms, but that specific peripheral relationships come apart, while the Bitcoin and Ethereum core stays intact in both tails.
 
Taken together, the two tails describe two different markets. The lower tail is a market that crashes as a single block throughout the sample, its near-complete dependence stable across cycles and its only periphery a single small asset. The upper tail is a market that rallies in an increasingly differentiated way, thinning and beginning to re-form sectoral structure as the sample progresses. Crashes draw the assets together; rallies, increasingly, do not.

\subsection{The asymmetry is statistically significant}\label{sec:asym}

Section~\ref{sec:42} established that lower-tail dependence exceeds upper-tail dependence in
every window. Because the windows overlap by some $98\%$, treating them as independent
observations would be invalid, so we test within individual windows instead. For each of three
representative windows, dated mid-2022, mid-2023 and mid-2024, we draw $300$ moving-block
bootstrap resamples of the filtered residuals, re-estimate both tails on each, and record the
difference between the panel-averaged lower- and upper-tail extremal correlations together with
the Bitcoin-Ethereum and the Tron-periphery correlations.

Table~\ref{tab:asymmetry} reports the resulting $95\%$ intervals. The gap between the lower and
upper tails is positive and excludes zero in every window, confirming that joint losses are
significantly more dependent than joint gains, and the interval for the Tron pairs lies
entirely below that for the core pairs in every window, confirming that the periphery is a
stable feature and not an artefact of a single estimate. The gap is, if anything, slightly
wider early in the sample, which locates the \emph{emergent} part of the asymmetry in the
upper tail's loss of connectivity over time rather than in any widening of the pairwise gap.

\begin{table*}[t]
\centering
\small
\caption{Within-window moving-block bootstrap ($300$ resamples). Each cell reports the median
and, in brackets, the $95\%$ interval. The first row is the difference between the
panel-averaged lower- and upper-tail extremal correlations; the remaining rows are the
Bitcoin-Ethereum correlation by tail and the lower-tail correlation of the Tron pairs against
the core pairs.}
\label{tab:asymmetry}
\begin{tabular}{@{}lccc@{}}
\toprule
 & Mid-2022 & Mid-2023 & Mid-2024 \\
\midrule
$\bar{\chi}^{-}-\bar{\chi}^{+}$
 & 0.11\,[0.09,\,0.12] & 0.08\,[0.07,\,0.10] & 0.08\,[0.06,\,0.10] \\
$\chi_{\text{BE}}$ lower
 & 0.80\,[0.78,\,0.82] & 0.78\,[0.75,\,0.81] & 0.76\,[0.73,\,0.78] \\
$\chi_{\text{BE}}$ upper
 & 0.74\,[0.71,\,0.77] & 0.73\,[0.70,\,0.76] & 0.71\,[0.69,\,0.73] \\
Tron pairs (lower)
 & 0.70\,[0.68,\,0.73] & 0.63\,[0.60,\,0.66] & 0.61\,[0.58,\,0.63] \\
Core pairs (lower)
 & 0.75\,[0.73,\,0.76] & 0.71\,[0.70,\,0.73] & 0.71\,[0.70,\,0.72] \\
\bottomrule
\end{tabular}
\end{table*}
\begin{figure}
\centering
\includegraphics[width=0.99\linewidth]{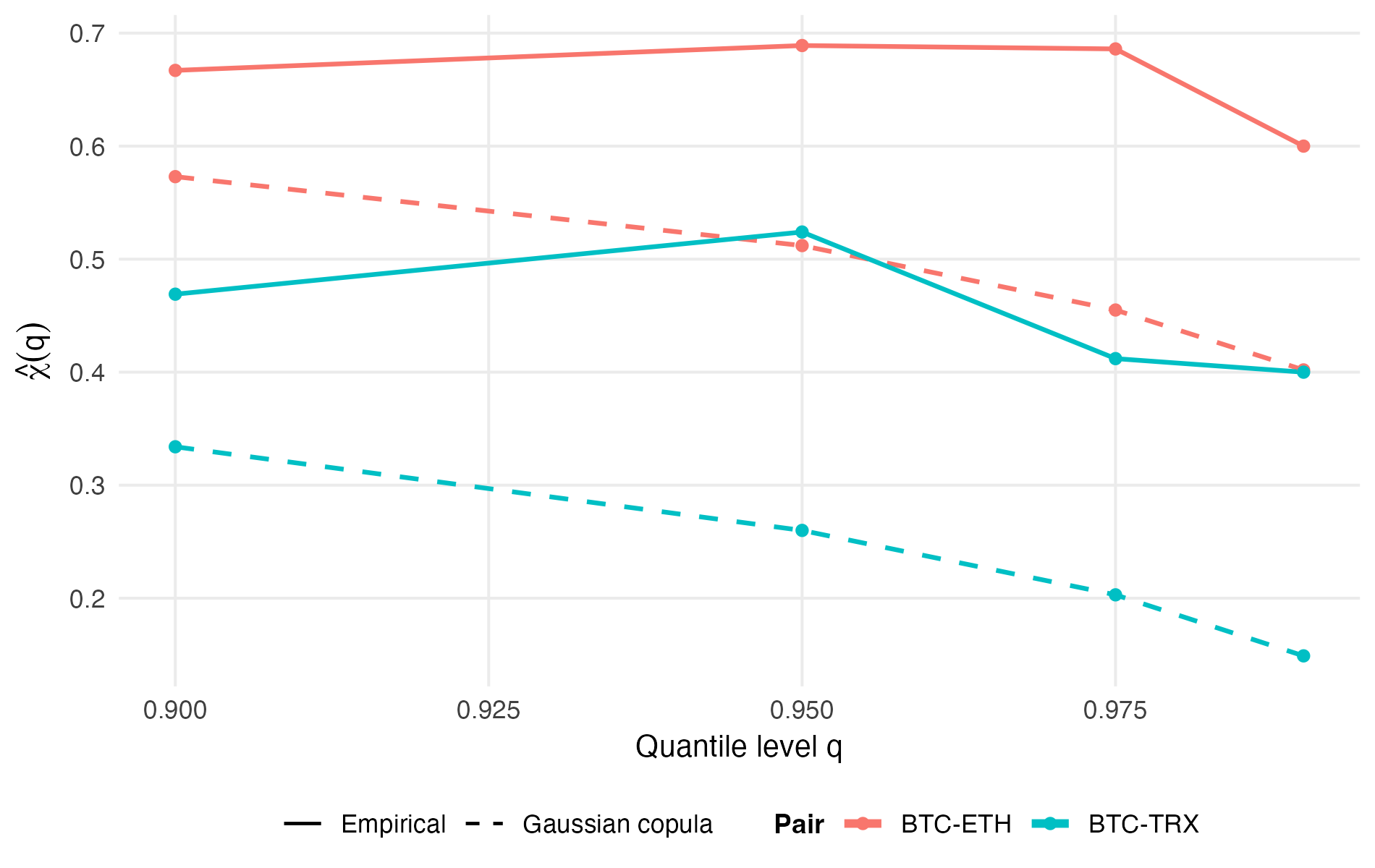}
\caption{Extremal correlation as a function of the quantile level for two pairs in the lower
(crash) tail. Solid lines are the empirical values; dashed lines are those implied by a
Gaussian copula with the same correlations, which decay toward zero.}
\label{fig:chiq}
\end{figure}

\subsection{The cost of the tail: how far covariance understates joint crashes}\label{sec:econ}
\label{sec:43}
The preceding results describe the shape of extremal dependence the HR model
recovers. We now ask whether that dependence is economically material, by turning it into the
probability that the market crashes as a whole, and whether the asymptotic dependence the model
encodes is borne out in the data against the natural alternative of a model that has none. The
foil is a Gaussian copula whose correlation matrix is the Pearson correlation of the
standardised residuals over the full sample, from which we simulate $200{,}000$ days; the
empirical quantities are computed on all $2{,}076$ days. So that nothing turns on the marginals,
every calculation below holds them fixed at their empirical values and varies only the
dependence; the comparison is also free of any graph, penalty or sparsity rule, so it is
untouched by the sensitivity documented in Appendix~\ref{app:robustness}.

The defining property of the HR model is asymptotic dependence: an extremal
correlation that stays positive no matter how deep into the tail one looks. A Gaussian
dependence structure has the opposite property, asymptotic independence, under which the
probability that one asset is extreme given that another is falls to zero as the level rises.
Figure~\ref{fig:chiq} adjudicates between the two. The empirical extremal correlation between
Bitcoin and Ethereum stays near $0.65$ across quantile levels from the $90$th to the $99$th
percentile, and that between Bitcoin and Tron near $0.45$, flat in the level just as asymptotic
dependence requires; the values implied by a Gaussian copula with the same correlations instead
decline steadily, from $0.57$ to $0.40$ for the first pair and from $0.33$ to $0.15$ for the
second. The flat empirical curves are the finite-level counterpart of the positive limiting
correlations the fitted model reports, $0.78$ for Bitcoin-Ethereum and lower but still positive
elsewhere (Section~\ref{sec:41}), which the curves need not reach at any finite $q$; what
matters is that they do not decay. The deeper into the tail one looks, the further the Gaussian
falls behind the data and the better the HR description fits.

\begin{figure}
\centering
\includegraphics[width=0.99\linewidth]{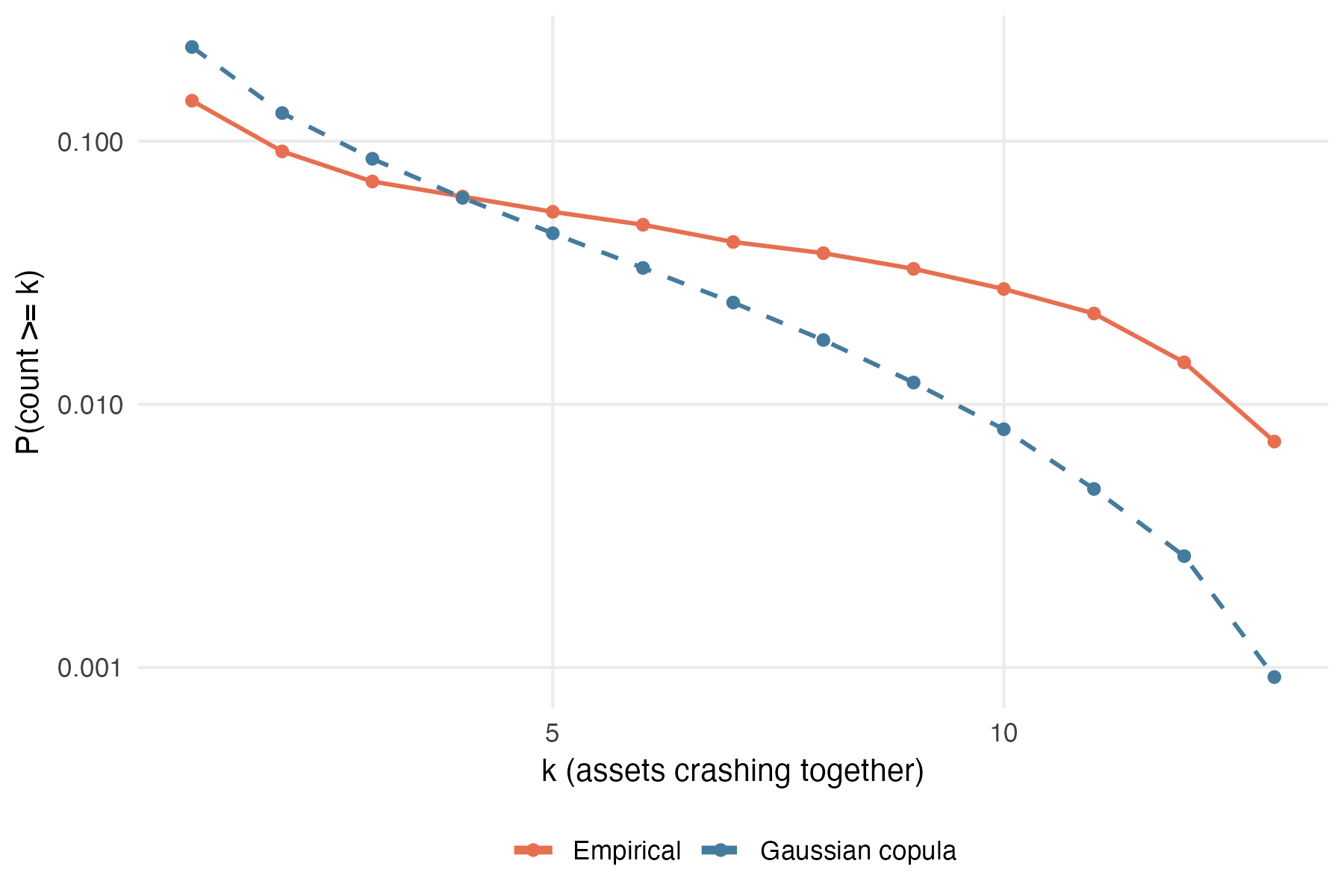}
\caption{Probability that $k$ or more of the thirteen assets breach their own lower $5\%$
quantile on the same day, empirically and under a Gaussian copula with the same correlations.
The vertical axis is on a log scale.}
\label{fig:joint}
\end{figure}

This asymptotic dependence has a direct multivariate cost. Because extreme moves do not
decouple as the level rises, the probability that the whole market crashes together does not
vanish the way a Gaussian model requires. Figure~\ref{fig:joint} reports the probability that
$k$ or more of the thirteen assets breach their own $5\%$ loss quantile on the same day,
empirically and under the Gaussian copula. For small $k$ the Gaussian probability is the higher
of the two, spreading risk into events in which a handful of assets fall together; the curves
cross at around four assets, and beyond that the empirical probability pulls away. The
probability that eight or more of the thirteen fall together on a $5\%$ day is $0.038$ in the
data against $0.018$ under the Gaussian, and the probability that all thirteen fall together is
$0.0072$ against $0.0009$, an understatement of roughly eightfold (about thirtyfold at the $1\%$
level, though that deepest figure rests on only three days in the sample). The Gaussian model
does not merely understate the magnitude of joint crash risk; it misplaces it, treating as a
scattering of partial co-movements what is in fact a tendency for the whole market to fall at
once.

The portfolio consequences are milder, because aggregation dilutes dependence.
Table~\ref{tab:portfolio} reports the value at risk and expected shortfall of an equal-weight
portfolio under the empirical dependence, a Gaussian copula, and independence, all sharing the
empirical margins. The $99\%$ expected shortfall is $3.19$ under the empirical dependence
against $2.65$ under the Gaussian, an understatement of about a fifth, though both lie far above
the $0.75$ that independence would imply. A covariance model is, in this precise sense, an
adequate guide to the typical loss of a diversified portfolio and a misleading guide to the
systemic event in which the whole market falls together. 

\begin{table}[t]
\centering
\small
\caption{Value at risk and expected shortfall of an equal-weight portfolio of the thirteen
assets, in units of the standardised return, under three dependence structures sharing the
empirical margins.}
\label{tab:portfolio}
\begin{tabular}{@{}lcccc@{}}
\toprule
Dependence & VaR$_{95}$ & ES$_{95}$ & VaR$_{99}$ & ES$_{99}$ \\
\midrule
Empirical        & 1.31 & 2.03 & 2.43 & 3.19 \\
Gaussian copula  & 1.22 & 1.75 & 2.05 & 2.65 \\
Independence     & 0.43 & 0.56 & 0.64 & 0.75 \\
\bottomrule
\end{tabular}
\end{table}




\section{Discussion}\label{sec:discussion}

The results in Section~\ref{sec:results} describe the conditional tail dependence structure of
the cryptocurrency market over four years of overlapping windows and, taken together, recast
several conclusions from the existing literature. We discuss four themes in turn: the failure
of diversification at the extremes, the time-varying asymmetry between crash and rally
dependence, the core-periphery structure of the market and the dissolution of token categories
under stress, and the implications for portfolio management, risk measurement and regulation.
We then set out the limitations of the analysis and the extensions they invite.

\subsection{Diversification fails at the extremes}

The most basic finding is that the network of tail dependence is structurally distinct from
the network of ordinary dependence: the extremal graphs are close to complete, with an edge
density near $0.87$ and a diameter of two, so that almost every pair of assets remains directly
dependent at the extremes. This extends the quantile connectedness evidence of
\citet{bouri2021quantile}, who document that connectedness among crypto-assets is higher in the
tails than at the mean. Where a connectedness measure records
the volume of spillover, our conditional analysis identifies which dependences survive once
the rest of the market is accounted for, and the answer is that almost all of them do: at the
extremes the conditional independences that organise ordinary co-movement very nearly vanish,
so that an extreme move in one asset is transmitted, directly rather than through
intermediaries, to almost every other.

For an investor this is the formal counterpart of the practitioner observation that
correlations rise towards one in a crisis, but it is a stronger statement. A rise in pairwise
correlation can in principle reflect a common factor that diversification across many assets
would still dilute; the near-completeness of the conditional graph rules this out, because it
shows that pairs remain directly dependent at the extremes even after conditioning on every
other asset. There is, in consequence, no internal hedge: holding more crypto-assets does not
reduce the probability of a joint loss, because the assets do not become conditionally
independent at the point where independence would matter. The contrast with the Gaussian
benchmark sharpens the warning, but not because the benchmark is sparser, since its density
turns on the estimator (Appendix~\ref{app:robustness}). It is sharper than that: a Gaussian
dependence model carries no asymptotic tail dependence, so the probability it assigns to many
assets crashing together falls away far faster than the data require. A risk model calibrated on
the covariance matrix, whether through mean-variance optimisation, a beta, or a
partial-correlation network, understates the probability of a market-wide crash many times over,
by roughly eightfold at the $5\%$ level and by an order of magnitude more in the extreme tail
(Section~\ref{sec:econ}), and reports a diversification that does not exist when it is needed.
This is the multivariate, conditional expression of the
heavy-tailed marginal risk documented by \citet{tzagkarakis2023} and \citet{osterrieder2016}.

\subsection{The crash-rally asymmetry, and why the literature disagreed}

The central dynamic finding is that the relative strength of crash and rally dependence is not
a fixed property of the market but a moving one. Lower-tail dependence exceeds upper-tail
dependence in every window and for every Bitcoin- or Ethereum-partner pair, by $0.08$ on
average, yet this ordering is recent: early in the sample the two tails are equally connected
and the upper tail is marginally the denser, and the gap then opens as the upper tail thins,
from a peak density of $0.92$ in 2022 to $0.81$ in 2024, while the lower tail holds near
$0.88$ throughout. Crashes draw the market into a single block across the whole sample;
rallies increasingly do not.

This time-variation offers a direct explanation for the long-standing and unresolved
disagreement in the bivariate literature over which tail dominates.
\citet{gkillas2018extreme} find lower-tail dependence the stronger, \citet{tiwari2020empirical}
and one of the two return-volume studies of Naeem and coauthors find the upper tail the
stronger \citep{naeem2020,naeem2020tail}, and \citet{chan2022extreme} find the link weakening
in the extreme tails. If the direction of the asymmetry shifts over time, as our windowed
estimate shows it does, then studies estimated on different sample periods will reach opposite
conclusions about a quantity that has no single time-invariant value. The upper-denser regime
we observe early in the sample is consistent with the upper-tail-dominant findings, and the
lower-denser regime that follows is consistent with the lower-tail-dominant findings; the
disagreement is then not a contradiction to be resolved in favour of one camp but an artefact
of treating a moving target as a fixed one. This is the multivariate, whole-market counterpart
of \citet{gong2022asymmetric}, who find the Bitcoin-Ethereum lower tail moving toward
dependence over time, and it reframes their result: what moves is not the lower tail, which is
near-complete throughout, but the upper tail, which thins.

The same finding reconciles the dispute over whether extremal dependence is structural or
time-varying. \citet{jlassi2023subsample} conclude that the crypto tail copula is
time-invariant and therefore structural, while \citet{de2020tracking} and
\citet{gong2022asymmetric} find it changing. Our results suggest both are partly right, but
about different parts of the tail. The lower tail is structural in the sense of
\citet{jlassi2023subsample}, near-complete and stable across cycles, whereas the asymmetry and
the upper-tail topology evolve in the sense of \citet{gong2022asymmetric} and
\citet{de2020tracking}. The financial reading is unfavourable to the diversified investor: the
diversification that survives is concentrated in the upper tail, where it is least valuable,
and it is absent in the lower tail, where it is most needed. Whatever protection the structure
of the market offers, it offers on the way up.

\subsection{The core, the periphery, and the dissolution of token categories}

Beneath the near-complete topology the market has a clear and stable hierarchy. Bitcoin and
Ethereum form the strongest conditional pair in both tails and anchor a dense core to which
the remaining large-capitalisation assets attach closely, while Tron sits consistently at the
periphery with the weakest extremal correlation to the core in either tail. That Bitcoin and
Ethereum are central is itself familiar, recovered as a two-cluster structure by
\citet{ahelegbey2021tail}, as net transmitters by \citet{bouri2021quantile}, and as systemic
hubs by \citet{naifar2025}. Our conditional analysis sharpens the claim: Bitcoin and Ethereum
are not merely high in degree on a graph where almost every asset is, they are the strongest
\emph{conditional} pair, so their linkage is direct and not an artefact of common exposure to
the rest of the market. Tron, conversely, is the one asset whose extremal dependence is
partly mediated by the others, which makes it the closest thing to an internal diversifier the
market contains, though with a lower-tail extremal correlation still near $0.65$ even that role
is modest.

The more consequential structural finding is that the functional categories into which
crypto-assets are conventionally sorted, currencies, smart-contract platforms, payment tokens
and infrastructure tokens, dissolve at the extremes. The Gaussian benchmark retains a sectoral
modularity around $0.15$, but both extremal graphs have modularity below $0.02$, so the
category boundaries that organise ordinary co-movement carry essentially no information about
joint crashes. This qualifies the increasingly common practice, exemplified by
\citet{naifar2025}, of organising crypto systemic-risk analysis by token type. The categories
are informative at the centre of the distribution and, as the slow rise in upper-tail
modularity shows, are beginning to re-form among joint gains; but in a crash they are
irrelevant, and a portfolio diversified across token categories is not diversified against
downside co-movement. The variogram clusters make the same point at the level of individual
pairs: the crash-time periphery is a single asset, whereas the rally-time periphery widens to
twenty-three pairs as specific links among the infrastructure, payment and smart-contract
tokens decouple, the disaggregated counterpart of the upper tail's thinning.

The shared arc of the estimates, high dependence through 2022, a trough around mid-2024 in the
period following the spot exchange-traded fund approval, and a recovery into the subsequent
rally, locates these structural features in market time. We read the trough cautiously: it
coincides with the arrival of exchange-traded-fund flows and a changing participant base, and
a loosening of internal dependence is a plausible accompaniment to such a broadening of
ownership, but the windowed design identifies the timing rather than the cause, and we do not
claim a causal channel.

\subsection{Implications for portfolio management, risk and regulation}

Three practical implications follow. First, for portfolio construction, the near-complete
lower-tail network implies that diversification within the crypto-asset class provides little
protection against joint losses. Spreading a position across more coins, or across token
categories, does not lower the probability of a simultaneous crash, because the assets are
directly conditionally dependent at the extremes regardless of category. The differentiation
that does exist sits in the upper tail and has grown over the sample, so asset selection can
shape upside participation but not downside exposure. An investor seeking genuine protection
against crypto crash risk must look outside the asset class, since the within-market hedge that
the sectoral structure appears to offer in normal conditions is not present when it is needed.

Second, for risk measurement, models built on average dependence are systematically
miscalibrated for the tails. A covariance matrix, a correlation-based diversification measure,
or a Gaussian and partial-correlation network carries no asymptotic tail dependence, and the
sectoral structure it does encode disappears under stress, so value-at-risk, expected shortfall and
stress scenarios calibrated on such objects understate the probability and the breadth of
joint crashes: in our data the chance that the whole market falls together on a given day is
understated roughly eightfold at the $5\%$ level, and by more than thirtyfold deeper in the tail
(Section~\ref{sec:econ}). This is the practical content of the distinction drawn in
Section~\ref{sec:lit} between the partial-correlation networks of \citet{naifar2025}, which
are a second-moment object, and a model of dependence in the tail itself: the two coincide in
ordinary conditions and diverge precisely in the regime that risk management exists to
address. Calibrating tail-risk tools on extremal dependence, rather than on average dependence
evaluated at extreme quantiles, would remove this bias.

Third, for systemic-risk monitoring and regulation, the analysis identifies both a locus and a
signal. The locus is the Bitcoin and Ethereum core, through which crash dependence is
transmitted directly and which is the natural focus of any tail-sensitive monitoring of the
market, consistent with the supervisory emphasis of \citet{naifar2025}. The signal is the
dynamic structure itself: the emergent asymmetry and the loosening and re-tightening of
dependence around major events are observable in close to real time from the windowed
estimates, which makes the dynamic extremal graph a candidate diagnostic for the kind of
tail-sensitive surveillance that the systemic-risk literature has called for. A regulator
organising oversight by token category should, however, note that those categories carry no
information about crash propagation, so a framework that monitors stablecoins, infrastructure
tokens and the rest as separate compartments will misread the way distress actually travels in
a downturn.

\subsection{Limitations and future research}

Three classes of limitation should be acknowledged. The first concerns the data and the choice
of assets. The windowed design requires a balanced panel, so the thirteen assets are those with
a common history across the whole sample, and the casualties of the period, most importantly the
Terra ecosystem tokens and the assets most damaged by the collapse of major exchanges, are
excluded by construction. These are precisely the assets through which the sharpest contagion of
2022 travelled, so the recovered structure is that of the survivors and, if anything,
understates the true completeness of crash dependence; the bias is therefore conservative for
our headline finding but real, and a design that admitted entering and exiting assets would
describe the crash network more faithfully. Stablecoins are excluded for a separate reason: a
pegged series has no meaningful continuous variation, so the marginal filtering on which the
dependence analysis rests does not apply to it, and what extremes it has are isolated
de-pegging events rather than the joint market moves we study. Whether a stablecoin acts as a
genuine safe haven or simply decouples at the extremes is an economically important question,
but one that calls for a different treatment of the margins and is left to future work.

The second concerns scope. The analysis is deliberately descriptive: it establishes the
conditional topology of the market in the tails and how that topology moves, but it does not
attempt to explain that movement through macroeconomic drivers. Mapping the structural shifts
we document, the mid-2024 trough and the emergent asymmetry, onto variables such as equity
volatility or the interest-rate cycle is a natural next step, but the heavy autocorrelation
induced by the overlapping windows precludes a standard time-series regression and calls for a
dedicated econometric framework, which we leave to subsequent work. The analysis is also
internal to the crypto market and conducted at daily frequency: it says nothing about tail
dependence between crypto-assets and equities, bonds or commodities, which the downside-spillover
evidence of \citet{hanif2022nonlinear} shows to be material, and it cannot see intraday
propagation. Extending the extremal graph to a joint crypto-and-traditional system, and to
higher frequency, are natural directions.

The third is methodological. The model is fitted rather than used to forecast; embedding the
dynamic extremal graph in a forecasting framework would turn the monitoring signal described
above into an operational early-warning tool. The windows are estimated independently, which
discards information shared across adjacent periods, and a hierarchical specification that let
neighbouring windows borrow strength would sharpen the trajectories, especially where the
sample is thin. And while we confirm in Section~\ref{sec:asym} that the lower tail is significantly more dependent
than the upper, the dating of the flip between the two regimes is left informal here;
changepoint methods for extremal dependence, developed in the lower-dimensional setting
\citep{de2020tracking,lattanzi2021}, could be extended to the graphical model to date the structural change
from the data. The robustness of the findings to the exceedance threshold, the window length
and the sparsity penalty, together with a like-for-like Gaussian benchmark, is examined in
Appendix~\ref{app:robustness}.



\section{Conclusion}\label{sec:conclusion}

This paper has applied HR graphical models of extremes to the thirteen largest
cryptocurrencies, estimated dynamically over a sequence of overlapping windows and separately
for joint crashes and joint rallies, and set against a GGM of ordinary
co-movement on the same data. The analysis characterises the conditional dependence structure
of the market in the tails of its return distribution, and how that structure has evolved as
the market has moved through the cycles of recent years.

Three findings stand out. The tail dependence structure is close to complete, so that
diversification within the asset class very nearly disappears at the extremes: in a crash almost
every pair of coins remains directly dependent once the rest of the market is accounted for, and
the sectoral structure that organises ordinary co-movement is absent. This near-completeness is stable in the lower tail
across the whole sample, whereas the upper tail thins over time and begins to re-form a faint
sector structure, so that the market crashes together throughout while rallying in an
increasingly differentiated way. The asymmetry between the two tails is therefore not a fixed
property of the market but one that has emerged over the sample, which explains why earlier
bivariate studies estimated on different periods disagreed over its direction. Beneath this
topology the market has a stable hierarchy, with Bitcoin and Ethereum forming the strongest
conditional pair in both tails and a single asset, Tron, at the periphery, while the token
categories dissolve in crashes.

Taken together, these results describe a market with two different architectures of
dependence: one for crashes, in which the assets behave as a single tightly connected block,
and one for rallies, in which they increasingly come apart. The practical consequence is that
the diversification crypto appears to offer in normal conditions is least available precisely
where it matters, on the downside, and that risk measures and diversification tools built on
the covariance matrix, which carry no asymptotic tail dependence and so understate the
probability of a market-wide crash many times over, are a systematically misleading guide to
joint crash risk. The dynamic extremal graph offers a more faithful tool, both for gauging diversification
tail by tail and for monitoring the concentration of systemic linkage in close to real time.

As the cryptocurrency market continues to institutionalise, through exchange-traded products
and the broadening participation that accompanies them, the structure of its extreme
co-movements will remain central to how investors diversify and how regulators monitor systemic
risk, and tail dependence rather than average dependence should be the primary lens for both.
Several extensions would sharpen the picture developed here: embedding the dynamic graph in a
forecasting framework, to turn the monitoring signal into an early-warning tool; extending the
system to include equities, bonds and commodities, so that the diversification question can be
answered across asset classes and not only within crypto; and dating the structural change in
the asymmetry formally rather than by inspection. Each builds naturally on the framework set
out here, and together they would carry it from a description of how the market has crashed and
rallied to a tool for anticipating how it will.

\printcredits

\bibliographystyle{cas-model2-names}

\bibliography{cas-refs}


\appendix

\section{Marginal filtering diagnostics}\label{app:diagnostics}

This appendix reports the diagnostics for the marginal filtering described in
Section~\ref{sec:filtering}. Each return series is filtered with a first-order
autoregressive mean and an asymmetric GJR-GARCH$(1,1)$ variance with Student-$t$
innovations, and the dependence analysis is conducted on the standardised residuals. We
check three things on every series: that the mean filter removes serial correlation, that the
variance filter removes volatility clustering, and that the residuals, once these dynamics are
gone, retain the heavy tails that an extremal model is designed to exploit.

\begin{figure*}[t]
\centering
\includegraphics[width=\linewidth]{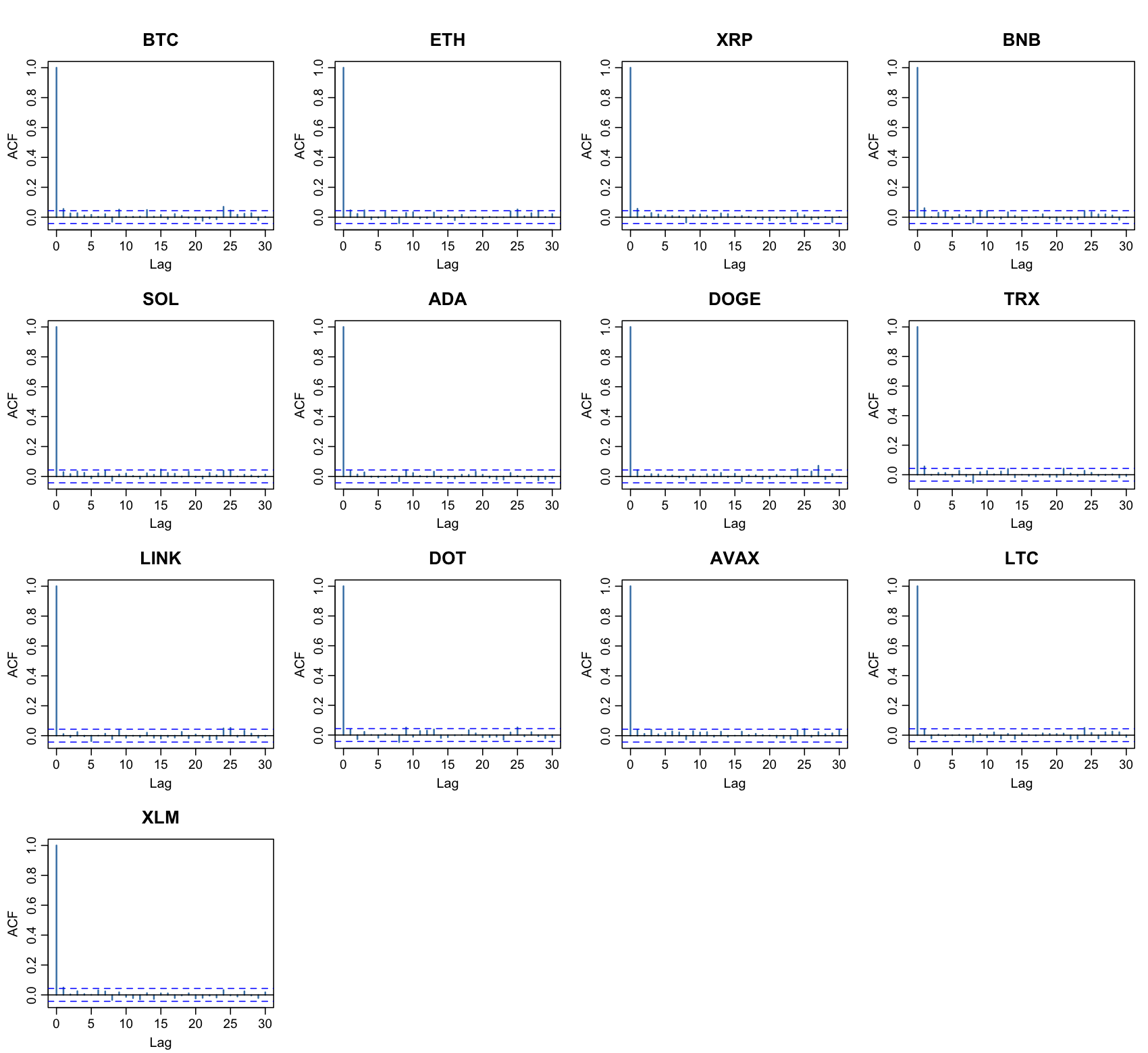}
\caption{Sample autocorrelation function of the standardised residuals, by asset. Dashed
lines mark the approximate $95\%$ confidence bands under no autocorrelation.}
\label{fig:acf_resid}
\end{figure*}

\begin{figure*}[t]
\centering
\includegraphics[width=\linewidth]{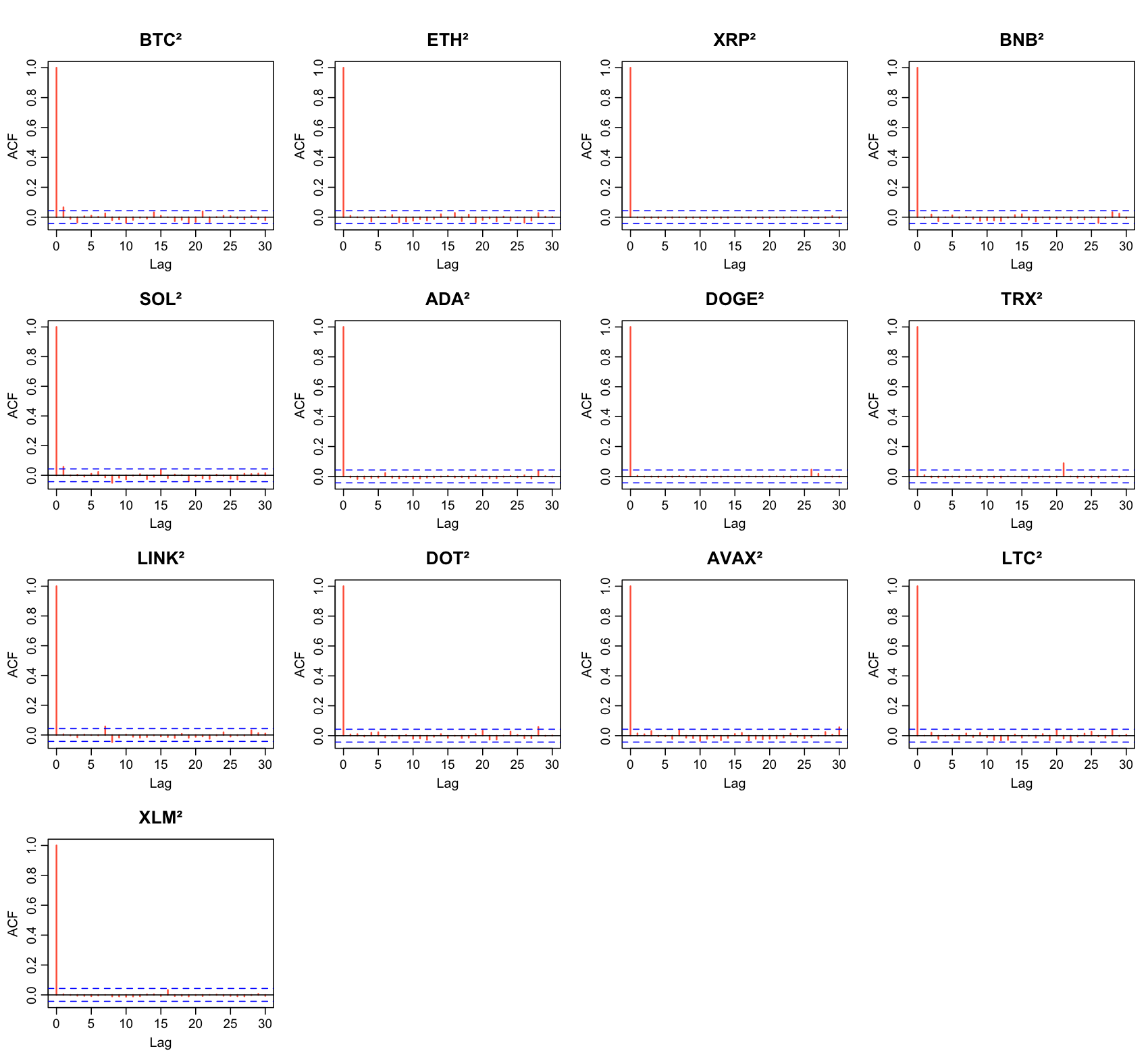}
\caption{Sample autocorrelation function of the squared standardised residuals, by asset.
Persistent autocorrelation here would indicate remaining volatility clustering; none is
present.}
\label{fig:acf_sq}
\end{figure*}

\begin{figure*}[t]
\centering
\includegraphics[width=\linewidth]{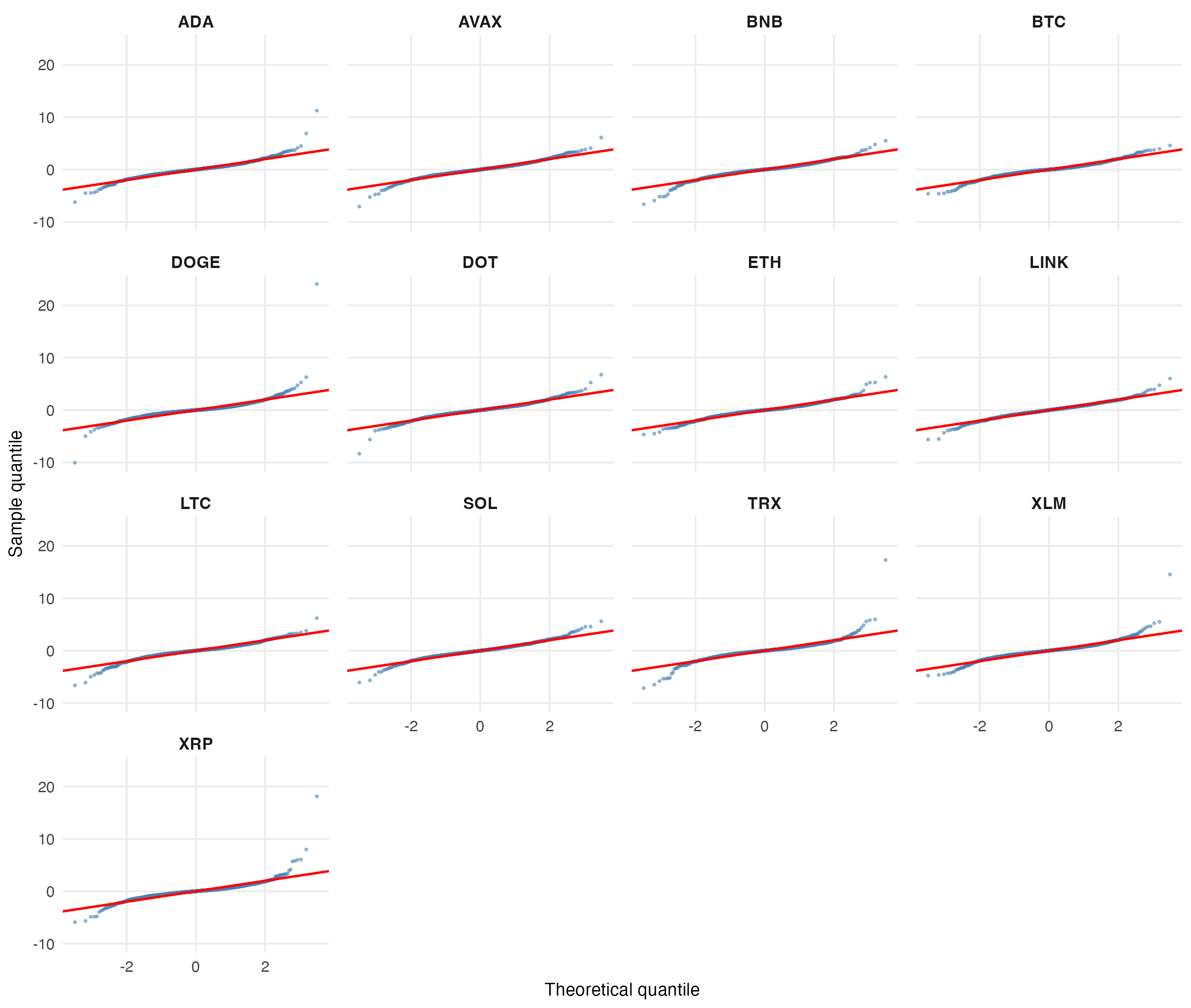}
\caption{Quantile-quantile plots of the standardised residuals against the standard normal,
by asset. Departure from the diagonal at both ends indicates heavy tails.}
\label{fig:qq}
\end{figure*}

\begin{figure*}[t]
\centering
\includegraphics[width=\linewidth]{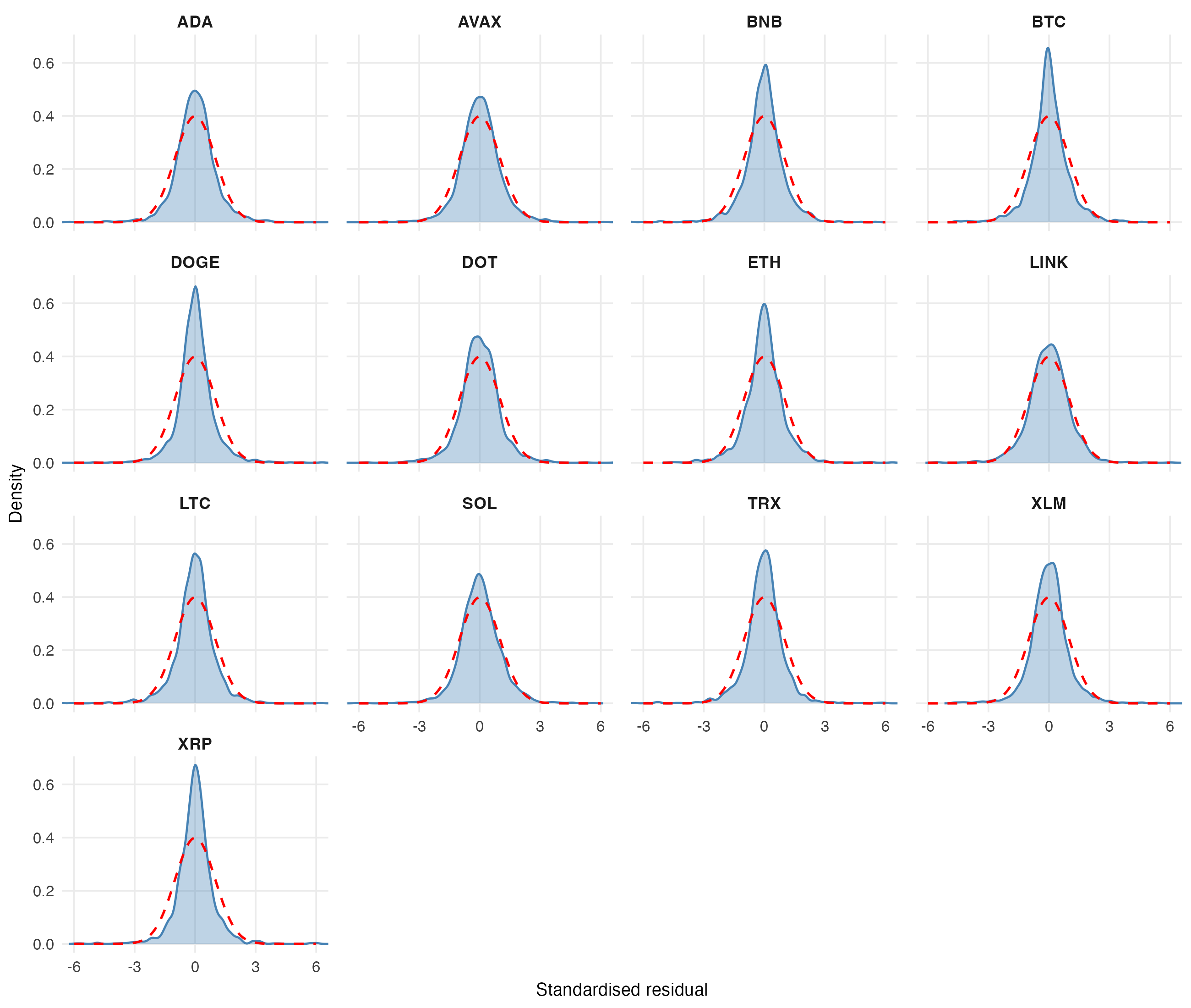}
\caption{Kernel density of the standardised residuals (solid) against the standard normal
density (dashed), by asset.}
\label{fig:density}
\end{figure*}

\begin{figure*}[t]
\centering
\includegraphics[width=\linewidth]{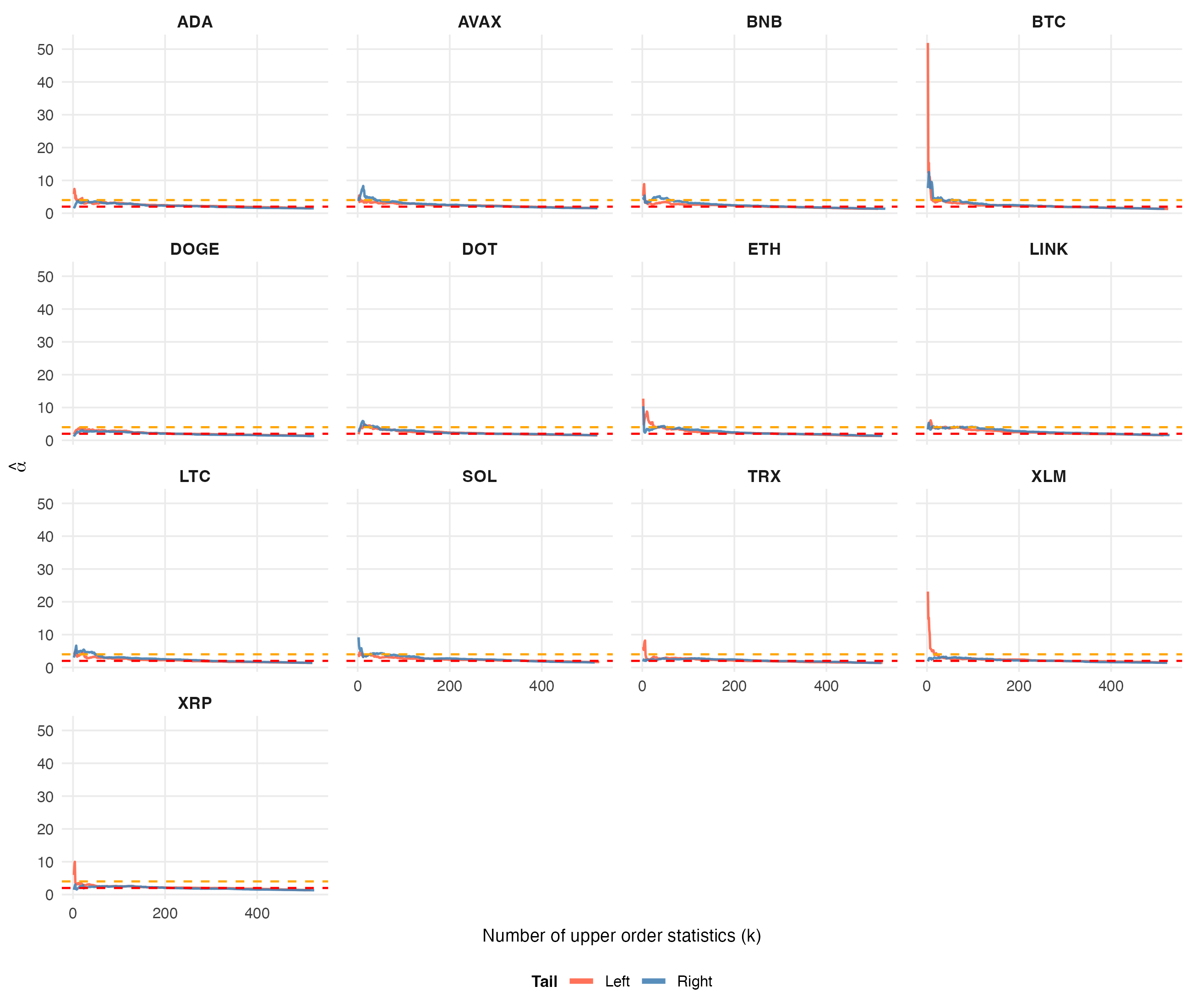}
\caption{Hill estimates of the tail index $\hat{\alpha}$ against the number of upper order
statistics $k$, by asset, for the left and right tails. Dashed lines mark $\alpha = 2$ and
$\alpha = 4$; estimates between them indicate finite variance but an infinite fourth moment.}
\label{fig:hill}
\end{figure*}

Table~\ref{tab:diagnostics} collects the numerical diagnostics and Figures~\ref{fig:acf_resid}
to~\ref{fig:hill} the graphical ones. The mean filter removes serial correlation from eleven
of the thirteen series; the two exceptions, BNB and ETH, show a Ljung-Box statistic that is
significant at the $5\%$ level, but the underlying autocorrelations are individually small and
lie close to the confidence bands (Figure~\ref{fig:acf_resid}). The variance filter is more
uniformly successful: the Ljung-Box test on the squared residuals is not rejected for any
asset, and the autocorrelations of the squared residuals lie within the bands at all lags
(Figure~\ref{fig:acf_sq}). The ARCH Lagrange-multiplier test is passed by all but Bitcoin,
where the rejection is marginal ($p = 0.033$) and the squared-residual autocorrelation is in
any case negligible.

What the filtering does not remove, and is not meant to remove, is the heaviness of the tails.
Normality is rejected for every asset by the Jarque-Bera test, the kernel densities are far
more peaked than the standard normal and carry heavier shoulders (Figure~\ref{fig:density}),
and the quantile plots bend away from the diagonal at both ends (Figure~\ref{fig:qq}). Excess
kurtosis remains large after filtering, exceeding one hundred for Dogecoin and falling below
three only for Chainlink. The Hill estimates of the tail index lie between $2.5$ and $3.6$
across the panel (Figure~\ref{fig:hill}), so the residuals retain finite variance but, for
most assets, an infinite fourth moment. Heavy tails of this order are precisely what a
covariance-based view of dependence would misrepresent, which motivates the extremal model and
its Gaussian benchmark in the main text.

\begin{table*}
\centering
\small
\caption{Diagnostics for the AR$(1)$--GJR-GARCH$(1,1)$-$t$ filtered residuals, by asset
($n = 2{,}076$). Excess kurtosis and the Hill tail-index estimate $\hat{\alpha}$ (upper
$10\%$) summarise the residual distribution; LB$(Z)$ and LB$(Z^2)$ are Ljung-Box tests at lag
$10$ on the residuals and on their squares; ARCH is the Lagrange-multiplier test with $5$
lags; JB is the Jarque-Bera test. The last four columns report $p$-values.}
\label{tab:diagnostics}

\begin{tabular}{@{}lrrrrrr@{}}
\toprule
Asset & Exc. kurt. & $\hat{\alpha}$ & LB$(Z)$ & LB$(Z^2)$ & ARCH & JB \\
\midrule
BTC  &   3.16 & 2.92 & 0.051 & 0.059 & 0.033 & $<$0.001 \\
ETH  &   3.78 & 3.00 & 0.026 & 0.672 & 0.853 & $<$0.001 \\
XRP  &  57.92 & 2.52 & 0.192 & 1.000 & 0.999 & $<$0.001 \\
BNB  &   4.94 & 2.87 & 0.010 & 0.900 & 0.772 & $<$0.001 \\
SOL  &   3.10 & 3.17 & 0.252 & 0.101 & 0.198 & $<$0.001 \\
ADA  &  10.56 & 2.98 & 0.284 & 0.961 & 0.902 & $<$0.001 \\
DOGE & 136.90 & 2.68 & 0.796 & 1.000 & 1.000 & $<$0.001 \\
TRX  &  43.00 & 2.52 & 0.060 & 1.000 & 0.995 & $<$0.001 \\
LINK &   2.56 & 3.55 & 0.507 & 0.233 & 0.991 & $<$0.001 \\
DOT  &   5.24 & 2.89 & 0.084 & 0.929 & 0.800 & $<$0.001 \\
AVAX &   3.68 & 2.91 & 0.161 & 0.399 & 0.773 & $<$0.001 \\
LTC  &   4.87 & 2.86 & 0.419 & 0.912 & 0.871 & $<$0.001 \\
XLM  &  24.07 & 2.61 & 0.219 & 1.000 & 0.996 & $<$0.001 \\
\bottomrule
\end{tabular}
\end{table*}

\section{Robustness and the role of the estimator}\label{app:robustness}

This appendix examines whether the findings of Section~\ref{sec:results} survive the main
modelling choices and clarifies what the
comparison with the Gaussian benchmark does and does not establish. Throughout we summarise
each network by its lower- and upper-tail edge density, its modularity, and the
Bitcoin-Ethereum extremal correlation, the quantities on which the substantive claims rest.

\subsection{Threshold and window length}

The exceedance threshold $p$ and the window length are the two choices most likely to drive
the results. We re-estimate the dynamic networks at $p = 0.10$, $0.15$ and $0.20$, holding the
window at $750$ days, and at window lengths of $500$, $750$ and $1000$ days, holding $p$ at
$0.20$; in each case the penalty is re-selected by cross-validation. We do not go below $p =
0.10$: at $p = 0.05$ a $750$-day window leaves on the order of forty joint exceedances with
which to fit a thirteen-node model, too few for stable estimation, and moderate thresholds are
in any case standard in graphical-extremes estimation in higher dimensions
\citep{engelke2020graphical}.

\begin{figure*}
\centering
\includegraphics[width=0.92\linewidth]{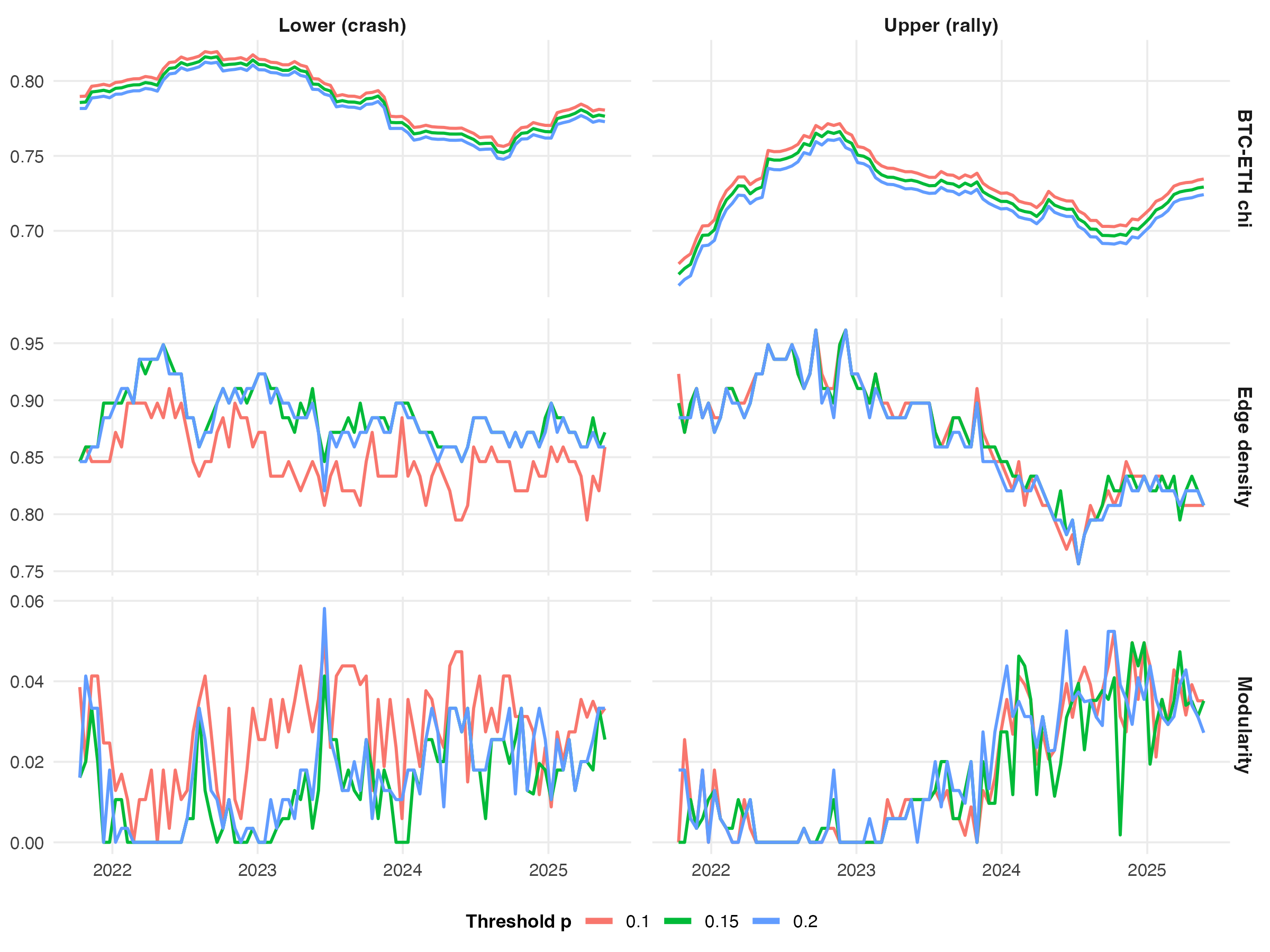}
\caption{Lower- and upper-tail edge density, modularity, and Bitcoin-Ethereum extremal
correlation over the windows, at exceedance thresholds $p = 0.10$, $0.15$ and $0.20$.}
\label{fig:rob_threshold}
\end{figure*}

\begin{figure*}
\centering
\includegraphics[width=0.92\linewidth]{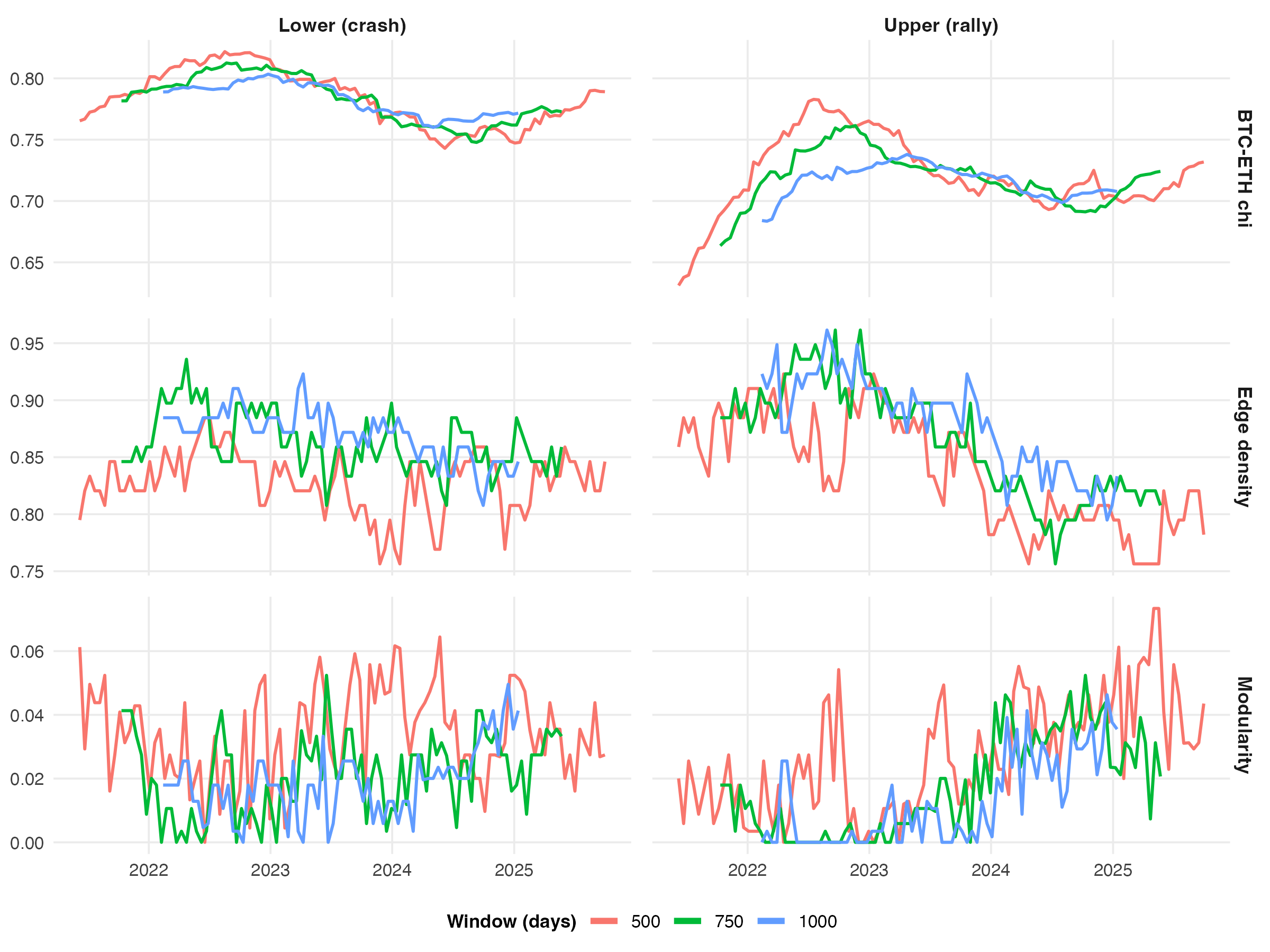}
\caption{The same statistics at window lengths of $500$, $750$ and $1000$ days, with the
exceedance threshold held at $p = 0.20$.}
\label{fig:rob_window}
\end{figure*}

Table~\ref{tab:robustness} reports the window-averaged statistics, and
Figures~\ref{fig:rob_threshold} and~\ref{fig:rob_window} the full trajectories. The headline
features are essentially invariant. The Bitcoin-Ethereum extremal correlation sits at $0.78$ in
the lower tail and $0.72$ in the upper across every threshold and every window length; the
lower tail remains near-complete while the upper tail thins; and the asymmetry between the two
holds throughout. The shorter $500$-day window is noisier and responds more sharply around the
market events, as expected, but tells the same structural story. The near-complete lower tail
and the emergent thinning of the upper tail are therefore properties of the market rather than
of the threshold or the window.

\begin{table*}
\centering
\small
\caption{Window-averaged network statistics under alternative exceedance thresholds (Panel A)
and window lengths (Panel B). Density and modularity are reported for the lower (crash) and
upper (rally) tails; $\chi^{-}_{\text{BE}}$ and $\chi^{+}_{\text{BE}}$ are the lower- and
upper-tail Bitcoin-Ethereum extremal correlations.}
\label{tab:robustness}
\begin{tabular}{@{}lrrrrrr@{}}
\toprule
 & \multicolumn{2}{c}{Density} & \multicolumn{2}{c}{Modularity} & \multicolumn{2}{c}{$\chi_{\text{BE}}$} \\
\cmidrule(lr){2-3}\cmidrule(lr){4-5}\cmidrule(lr){6-7}
Setting & Lower & Upper & Lower & Upper & Lower & Upper \\
\midrule
\multicolumn{7}{@{}l}{\textit{Panel A. Threshold $p$ (window $750$ days)}}\\
$p=0.10$ & 0.85 & 0.87 & 0.027 & 0.017 & 0.79 & 0.73 \\
$p=0.15$ & 0.89 & 0.87 & 0.014 & 0.016 & 0.79 & 0.73 \\
$p=0.20$ & 0.88 & 0.86 & 0.016 & 0.018 & 0.78 & 0.72 \\
\midrule
\multicolumn{7}{@{}l}{\textit{Panel B. Window length (threshold $p=0.20$)}}\\
$500$ days  & 0.83 & 0.84 & 0.034 & 0.027 & 0.78 & 0.72 \\
$750$ days  & 0.86 & 0.86 & 0.021 & 0.017 & 0.78 & 0.72 \\
$1000$ days & 0.87 & 0.88 & 0.019 & 0.012 & 0.78 & 0.72 \\
\bottomrule
\end{tabular}
\end{table*}

\subsection{Graph density and the choice of estimator}

The extremal graphs are near-complete, with an edge density close to $0.87$. The Gaussian
benchmark estimated as in Section~\ref{sec:ggm} is sparser, with a density near $0.46$, but
this gap should not be read as a difference between tail and ordinary dependence, because the
density of a graphical model is governed by the sparsity-selection rule as much as by the data.
Table~\ref{tab:benchmark} makes the point directly: a frequentist graphical lasso fitted to the
same standardised residuals, with its penalty chosen by an information criterion, returns a
Gaussian graph that is \emph{denser} than the extremal graphs. With $750$ observations and only
thirteen assets, almost every partial correlation is statistically detectable, so a Gaussian
graph can be made nearly complete or quite sparse depending on the rule applied.

We therefore do not rest the contrast between ordinary and extreme dependence on edge density.
The contrast that no choice of estimator can repair is the one reported in
Section~\ref{sec:econ}: a Gaussian dependence model, whatever its graph, carries no asymptotic
tail dependence, so the probability it assigns to many assets crashing together falls away far
faster than the data require. That is the substantive sense in which a covariance-based view of
the market understates joint crash risk.

\begin{table*}
\centering
\small
\caption{Window-averaged edge density of the extremal graphs and of two Gaussian graphs fitted
to the same residuals: the Bayesian model of Section~\ref{sec:ggm}, with edges included by
$95\%$ credible interval, and a frequentist graphical lasso with an information-criterion
penalty. Graph density is not a stable property of the dependence regime.}
\label{tab:benchmark}
\begin{tabular}{@{}lr@{}}
\toprule
Network & Edge density \\
\midrule
Lower-tail extremal (Hüsler-Reiss) & 0.86 \\
Upper-tail extremal (Hüsler-Reiss) & 0.86 \\
Gaussian, Bayesian ($95\%$ credible interval) & 0.46 \\
Gaussian, graphical lasso (information criterion) & 0.89 \\
\bottomrule
\end{tabular}
\end{table*}

\end{document}